*First-Principles Experimental Demonstration of Ferroelectricity in a Thermotropic Nematic Liquid Crystal:*
*Spontaneous Polar Domains and Striking Electro-Optics*

X. Chen[1], E. Korblova[2], D. Dong[3], X. Wei[3], R.F. Shao[1], L. Radzihovsky[1],
M.A. Glaser[1], J.E. Maclennan[1], D. Bedrov[3], D.M. Walba[2], N.A. Clark[1]

[1]*Department of Physics and Soft Materials Research Center,*
*University of Colorado, Boulder, CO 80309, USA*

[2]*Department of Chemistry and Soft Materials Research Center,*
*University of Colorado, Boulder, CO 80309, USA*

[3]*Department of Materials Science and Engineering, University of Utah, Salt Lake City, UT 84112, USA*
*and Soft Materials Research Center, University of Colorado, Boulder, CO 80309, USA*

*Abstract*

We report the experimental determination of the structure and response to applied electric field of the lower-temperature nematic phase of the previously reported calamitic compound 4-[(4-nitrophenoxy)carbonyl]phenyl2,4-dimethoxybenzoate (RM734). We exploit its electro-optics to visualize the appearance, in the absence of applied field, of a permanent electric polarization density, manifested as a spontaneously broken symmetry in distinct domains of opposite polar orientation. Polarization reversal is mediated by field-induced domain wall movement, making this phase ferroelectric, a 3D uniaxial nematic having a spontaneous, reorientable, polarization locally parallel to the director. This polarization density saturates at a low temperature value of ~ 6 $\mu C/cm^2$, the largest ever measured for a fluid or glassy material. This polarization is comparable to that of solid state ferroelectrics and is close to the average value obtained by assuming perfect, polar alignment of molecular dipoles in the nematic. We find a host of spectacular optical and hydrodynamic effects driven by ultra-low applied field (E~1 V/cm), produced by the coupling of the large polarization to nematic birefringence and flow. Electrostatic self-interaction of the polarization charge renders the transition from the nematic phase mean-field-like and weakly first-order, and controls the director field structure of the ferroelectric phase. Atomistic molecular dynamics simulation reveals short-range polar molecular interactions that favor ferroelectric ordering, including a tendency for head-to-tail association into polar, chain-like assemblies having polar lateral correlations. These results indicate a significant potential for transformative new nematic physics, chemistry and applications based on the enhanced understanding, development, and exploitation of molecular electrostatic interaction.

*<u>Significance</u>*

Conspicuously in the background in the history of liquid crystals is the ferroelectric nematic ($N_F$) phase. Predicted by Debye and Born a hundred years ago, and since revisited extensively, in systems ranging from colloidal suspensions of rods or discs to melts of polar molecules, the existence of the $N_F$ has never been certain, and it has never emerged in interest or applicability from the shadow of its familiar cousin, the dielectric nematic, the key component of the displays that enabled the portable computing revolution of the twentieth century. Here we show, in a previously reported thermotropic material, defining evidence for ferroelectricity and a host of novel polar behaviors that promise to remake the science and technology of nematics.

*Introduction*

The first theoretical treatments of collective molecular orientation in liquids, by Debye [1] and Born [2], were electrostatic versions of the Langevin-Weiss model of the paramagnetic/ferromagnetic transition in solids [3]. Born envisioned the orientational ordering of rod-shaped molecules of a nematic as a phase transition, the proposed ordering mechanism being the interaction of molecular electric dipoles, so that the resulting nematic phase was ferroelectric, i.e., predicted to have a spontaneous non-zero polarization density. Thus, the notion of LCs with polar order, introduced more than a century ago, has grown as a field of broad interest and challenge at the frontiers of LC science, stimulating rich themes of novel chemistry and physics [4,5,6 7,8,9,10,11,12,13]

However, following Born's model, some calamitic molecules without molecular dipoles were found to exhibit nematic phases [14], while ferroelectricity failed to materialize as a molecular nematic phenomenon. Born's calculation thus appeared to be incomplete, stimulating a variety of different models of nematic ordering in which both steric and/or electrostatic interactions were considered. These included the Maier-Saupe theory [15], where steric interactions produced apolar (quadrupolar) order, and others in which the nematic ordering could also be polar [4,5,16,17,18,19,20,21,22]. The appearance of polar ordering in these models and Born's is considered to be an equilibrium transition between bulk phases of higher and lower symmetry [23,24]. The models propose order parameters constructed to characterize this change of symmetry, and provide benchmarks for experimental testing, predicting pretransitional behavior as the phase transition is approached from higher or lower temperature, as well as describing the properties of the polar ordered phase and its distinct symmetry-related states. In the case of relevance here, of a uniaxial, non-polar nematic transitioning to a uniaxial, polar nematic with the polarization along the director, there are two ordered states related by reflection through a plane normal to the polarization. If such states coexist in a sample, they must form reflection-related domains with opposite polarization separated by well-defined domain walls [25,26,17]. Such polar domains and their boundaries are also described by the models, specifically by the elasticity and order parameter energetics of the polar phase, making the domains the signature features of spontaneous polar ordering to be probed and understood in characterizing the nature of the phase transition. If such domain boundaries can be moved or removed by application of a field then the mean polarization can be changed, and if this motion is irreversible, then the polar phase will exhibit switching and hysteresis as emergent properties and can be considered macroscopically ferroelectric [27]. Here we present the direct observation of such spontaneously broken symmetry in the form of domains of opposite polarization, grown without applied electric field, as a first-principles demonstration of ferroelectricity in a thermotropic, uniaxial, nematic liquid crystal of rod-shaped molecules.

In 2017 Mandle, et al. [28], and Kikuchi et al. [29] separately reported new liquid crystal (LC) compounds exhibiting unusual phase behavior: two distinct, fluid nematic phases separated in temperature by a weakly first-order phase transition. In both cases, the molecules were rod-shaped, with several intramolecular dipoles distributed along their length whose projections onto the molecular long axis summed to a large overall axial dipole moment of ~ 10 Debye. The high temperature phase of both mesogens was reported to be a typical nematic but they exhibited dramatic paraelectric [30,29] and ferroelastic [30] pretransitional effects, with a dielectric constant surpassing 1000 as the transition to the low temperature phase was approached. The low temperature phase exhibited enhanced dipolar molecular associations, reported to be antiparallel in the Mandle system [28], and suggested to be parallel in the Kikuchi system, the latter being termed "ferroelectric-like" [29], giving macroscopic polar ordering in response to an applied electric field. Mandle et al. subsequently synthesized a number of homologs of their molecule in an effort to develop structure-property relationships for this phase [31], and pursued, in collaboration with the Ljubljana group, a series of physical studies on one of these (RM734), shown in *Fig. 1A*, leading to the claim that this phase was locally polar, as evidenced by second harmonic generation, but, on some longer scale, an antiferroelecric "splay nematic" [32,33,30,34,35], a modulated phase stabilized by local director splay, of the type originally proposed by Hinshaw et al. [36].

Our resynthesis of RM734 (*SI Appendix Sec. S1*) and observation of its electro-optic behavior using polarized light microscopy provides no evidence for a splay nematic phase but rather leads us to the unambiguous conclusion that, upon cooling from the higher temperature, non-polar, uniaxial nematic (N) phase, RM734 undergoes a transition to another uniaxial nematic ($N_F$) phase that is ferroelectric. The key evidence for this result is the first observation in a nematic liquid crystal of the defining characteristics of ferroelectricity: (*i*) the formation, in the absence of applied electric field, of spontaneously polar domains of opposite sign of polarization separated by distinct domain boundaries; and (*ii*) field-induced polarization reversal mediated by movement of these domain boundaries, as summarized in *Fig. 1*.

In the N phase, the local texture of the planar-aligned cell shown in *Fig. 1* is optically featureless. On cooling toward the $N_F$ phase, a random pattern of stripes extended along the buffing direction appears. Once in the $N_F$ phase, these stripes coarsen, leading to a texture that is again local optically featureless (*Figs. 1B-D*) but characterized on a larger scale by a pattern of well-defined lines, some delineating distinct, lens-shaped domains 100 *μ*m or more in extent (*Figs. 1D-G*), all formed in the absence of applied electric field. Application of an ultra-small (~1 V/cm), in-plane, DC test field, *E*, applied along *z*, parallel to the in-plane buffing and therefore to the director *n*(*r*), shows that for $E > 0$ the director inside these domains begins to reorient while the orientation outside remains fixed (*Fig. 1E*), whereas for $E < 0$ the region outside the lens-shaped domains reorients and the orientation inside remains fixed (*Fig. 1G*), indicating that the domain boundaries separate regions with opposite response to in-plane field, and therefore of opposite in-plane polarization. The lack of response to increasing *E* outside of the lenses in *Fig. 1E* and inside the lens in *Fig. 1G* shows that *P* and *n* are co-linear in these domains in the field-free condition. Increasing the field causes the domain boundaries to unpin, shrink, and disappear (*Fig. 1F*), moving hysteretically to increase the area of the field-preferred state. The observations of *Fig. 1*, constituting a first-principles demonstration of nematic ferroelectricity, are described in more detail below and in *SI Appendix Secs. S3,S4.*

*Results and Discussion*

*Electro-Optic (EO) Observation of Planar-Aligned Cells* - Depolarized transmission light microscopy (DTLM) observations of RM734 were made principally in cells with a $t$ = 11 *μ*m-wide gap between the glass plates, one of which was coated with a pair of planar ITO electrodes uniformly spaced by $d$ = 1.04 mm, which enabled application of an in-plane electric field, *E*, between them, largely parallel to the cell plane (*y,z*). The plates were treated with a polyimide layer buffed in the *z* direction, normal to the electrode edges, so that the applied field was along the buffing direction: *E* = *zE* (*SI Appendix* **Fig. S3**). The cells were filled by capillarity with the liquid crystal in the isotropic phase. Both the N and $N_F$ phases were studied, with results as follows.

*Nematic Phase* – When cooled into the Nematic (N) phase, RM734 formed textures with the nematic director, *n*(*r*), the local mean molecular long axis and the optic axis, parallel to the plates (planar alignment). The white-light birefringence color at $T$ = 140°C was a uniform pale yellow-orange, in the third-order Michel-Levy band (retardance ~ 1500 nm) [37], with the larger index for optical polarization along *n*. The azimuthal orientation of *n*(*r*) was generally along *z*, but with a fixed pattern of in-plane orientational defects and weak continuous variations of the in-plane orientation, as seen in *SI Appendix* **Figs. S4,S6A**, suggestive of a relatively weak coupling to the azimuthal anisotropy of the surface. Measurements give a uniaxial birefringence $\Delta n$ ~ 0.2 (*SI Appendix* **Fig. S17**), suggesting that the alignment is planar, with *n*(*r*) nearly parallel to the plane of the plates. Tilting the cell away from being normal to the light beam did not reveal significant tilt of *n*(*r*) out of the cell plane but a small (few-degree) pretilt may be present. In addition to the locally uniform, planar texture imposed by the surfaces upon cooling into the N phase, we observed a few twisted areas (*SI Appendix* **Fig. S6A**), but generally the local preferred orientation was the same on the two plates and therefore likely established by a combination of the buffing with surface memory [38] of the nematic director pattern as it was first growing out of the isotropic.

*Ferroelectric Nematic Phase* - Upon cooling through the N–$N_F$ transition, the cell becomes patterned with an texture of irregular domains extended locally parallel to *n*(*r*), first appearing on a submicron scale but then annealing over a roughly 2°C interval into patterns of elongated lines of low optical contrast (***Fig. 1B***) that are also oriented generally along *n*(*r*). Some lines coarsen and extend along *n* while others form closed loops, 10–200 microns in extent, having a distinctive characteristic lens shape, elongated along *n*(*r*) (***Fig. 1 C,D***). Sample textures from this evolution are shown in ***Fig. 1***, with additional images in *SI Appendix* **Figs. S4-S9**. Upon completion of the transition, apart from these loops, the texture is smooth and very similar to that of the N phase at high *T* (*SI Appendix* **Fig. S6**). A small increase of *Δn* is observed at the transition (*SI Appendix* **Fig. S17**), and *Δn* then increases continuously with decreasing temperature, *T*, behavior in agreement with that previously reported for the nematic order parameter [33].

The $N_F$ phase exhibits striking electro-optic behavior in DTLM, starting with the response to the application of tiny (< 1 V/cm) in-plane electric fields. This sensitivity was exploited to probe and understand the static and dynamic changes of *n*(*r*) and *P*(*r*) in applied electric fields. Typical textures observed in these experiments, and the responses used to identify ferroelectric domains and determine their polarization orientation, are indicated in ***Fig. 2***. In the absence of applied field, the LC director *n* in these cells is generally along the buffing direction *z* and we observe domains separated by distinct boundaries. As in ***Fig. 1***, one set of domains responds only weakly to fields applied along *z*, indicating that in these regions *P* is already parallel to *E* and therefore that before the field is applied *P* is everywhere either nearly paral-lel to or antiparallel to *z*. In an applied field, in the domains where *P* is nearly antiparallel to *E*, the polarization responds by rotating toward *E*. The optical response to test fields makes the difference in polarity readily distinguishable, leaving no doubt that these domains are polar. The green vectors in *Fig. 2* indicate the field-induced reorientation of *P(r)* in the mid-plane of the cell. Tilting of the cell shows that this reorientation is azimuthal [$\varphi(r)$] about *x*, with *n(r)* remaining parallel to the (*y,z*) plane.

In an applied electric field, the *n(r)* field of the preferred states becomes only slightly better aligned along the field direction, evidence that *n* and *P* are parallel. We considered the possibility that the $N_F$ phase is biaxial, with *n* along *z* and secondary directors oriented preferentially parallel and normal to the cell surfaces. When such a biaxial phase grows in from a higher temperature uniaxial phase, the cell generally exhibits arrays of characteristic disclination lines where 180º flips of the secondary directors have become trapped at one or both surfaces [39]. If such structures were present in our cells, they would show up clearly in the polarized light microscope but none are observed. We conclude that the bulk $N_F$ condition is uniaxial, with *n* parallel to *P* in the absence of applied field. When an electric field is applied, *n(r)* experiences electric torques through its coupling with *P*, which we assume remains substantially locally parallel to *n*, *i.e.*, that the local optic axis is parallel to *P*.

*Fig. S4* shows a larger area of the $t = 11$ µm cell in orientations having the average director along the crossed polarizer/analyzer direction (A, F-H), and at 45º to it (B-E), illustrating that the overall textural alignment gives reasonably good extinction between crossed polarizers but with some brighter regions allowed because of the softness of the anchoring as noted above. The images show polarization reversal driven by an adjustable DC in-plane electric field. The textures of *n(r)* in the limiting states of plus or minus *E* are identically black but separated by a striking scenario of domain wall formation, coarsening and disappearance, all in the weak DC field range -2 V/cm < *E* < 2 V/cm. The field-aligned states extinguish between crossed polarizers as in (A), meaning that that they have *n(r)* everywhere parallel to z, and show a pink birefringence color in the third-order Michel-Levy band as in (B). The intermediate states have *n(r)* in the (*y,z*) plane but with spatial variation of its azimuthal orientation $\varphi(r)$ about *x*. This lowers the effective retardance of these regions, moving their birefringence down into the second- and first-order Michel-Levy bands and producing intense birefringence colors. The uniformly oriented domains obtained following field reversal are states in which the *n,P* couple has been reoriented in the bulk LC and also flipped on the aligning surfaces, the latter mediated by domain wall motion.

The field-induced reorientations in the planar-aligned geometry of *Figs.* **1** and **2** are twist deformations of the azimuthal orientation of *n(r)* about *x*, having the form $\varphi(r) = \varphi_c(x) \cos(\pi x/t)$ for small $\varphi_c$, where $\varphi_c(x)$ is the reorientation pattern in the cell mid-plane. This deformation can be generated in a uniaxial dielectric N phase using an in-plane AC electric field to induce a twist Freedericksz transition, for which the threshold field will be given by $E_D = (\pi/t)\sqrt{(K/\varepsilon_o \Delta\varepsilon)}$ [40]. Assuming a cell gap $t = 11$ µm, and typical nematic values of Frank elastic constant $K \sim 5$ pN, and a dielectric anisotropy $\Delta\varepsilon \sim 5$, one finds $E_D \sim 1000$ V/cm, giving an estimate which sets the field scale for typical in-plane dielectric nematic electro-optics. The fields required to produce the reorientations in the $N_F$ phase in *Figs. 1,2* are three orders of magnitude smaller. In small applied fields, electrical torque on the director field $\boldsymbol{\tau}_E = \boldsymbol{P} \times \boldsymbol{E}$ comes from the coupling of field to polarization. With this polar coupling and $P(x)$ starting antiparallel to $E$, our observed field-induced reorientations are polar Freedericksz transitions for which the torque balance equation is $K_T\varphi_{xx} + PE\sin\varphi(x) = 0$ [41,40]. This gives a field threshold of $E_P = (\pi/t)^2(K_T/P)$, and a cell midplane reorientation of $\varphi_c(E) \approx \sqrt{[6(E - E_P)/E_P]}$, so that a $\varphi_c = 0$ to $\varphi_c = 90º$ reorientation occurs in the field range $E_P < E_{0-90} < 1.4\ E_P$. Measurements of $E_{0-90}$ yield an experimental value of the threshold of $E_P \sim 1$ V/cm, from which we can estimate $P$. Taking $K_T$ to be in the range 2 pN $< K_T <$ 5 pN gives a value for $P$ in the range 3 $\mu$C/cm$^2 \lesssim P \lesssim$ 6 $\mu$C/cm$^2$.

We also measured $P(T)$ directly from the field-induced current associated with polarization reversal. We used both square- and triangle-wave driving fields in several different, two-terminal cell geometries, including: an in-plane cell similar to that used for the electro-optics but with gold electrodes (*SI Appendix* **Fig. S3**); 100 $\mu$m-thick sandwich cells with conventional gold electrodes; and a 0.5 mm diameter, cylindrical capillary with the LC in a 150 $\mu$m gap between planar electrode faces oriented normal to the cylinder axis (*SI Appendix*, **Fig. S3**). Time integration of the current signals obtained using these geometries produced consistent values of the polarization density as a function of temperature. Typical polarization data obtained from the in-plane cell using a square-wave drive frequency f = 200 Hz are shown in **Fig. 3**. The cell current showed only small capacitive and resistive contributions in the N phase, but, upon cooling into the N$_F$ phase, became dominated by an additional peak which exhibited the characteristics of ferroelectric LC polarization current: carrying a driving amplitude-independent net charge $Q$ (from which $P$ was calculated); and exhibiting a risetime $\tau$ varying inversely as the drive amplitude (**Fig. 3C**). The resulting $P$ increases continuously from small values at the transition, saturating at low $T$ at a value $P \sim$ 6 $\mu$C/cm$^2$, a value comparable to the ~4 $\mu$C/cm$^2$ found by Kikuchi et al. in the compound DIO [29]. Polarization in the N – N$_F$ transition temperature region may include pretransitional contributions in the N phase due to the divergence of $\varepsilon_{\parallel}$ [30], but this has not yet been studied in detail. The estimate above of $P$ from the polar twist threshold $E_P$ at $T = 120$ºC agrees well with the $P$ data in **Fig. 3**, indicating that the magnitude of spontaneous polarization achieved field-free in domains grown and cooled from the N phase is comparable to that induced by a field in the N$_F$ phase, as expected for a ferroelectric.

Further significance of $P \sim$ 6 $\mu$C/cm$^2$ can be appreciated by calculating a polarization estimate $P_e = p/v$, where $p$ = 11 Debye is the axial molecular dipole moment of RM734 [31] (*SI Appendix* **Secs. S9,S10**) and $v$ is the volume/molecule in the phase, $v$ = 325 cm$^3$/mole = 540 Å$^3$/molecule, assuming a LC mass density of $\rho$ = 1.3 g/cm$^3$ (*SI Appendix* **Sec. S9,10**). Using these parameter values and assuming complete polar ordering of the molecular long axes, we find $P_e \sim$ 6.8 $\mu$C/cm$^2$, comparable to our measured $P$ at low $T$ and indicating that RM734 has extremely strong spontaneous macroscopic polar ordering, a condition impossible in any state with domains of competing polarization as in the proposed splay nematic, for example. This value of $P$ is confirmed by our atomistic MD simulations, which yield $P \approx$ 6.2 $\mu$C/cm$^2$ in an equilibrated simulation of 384 molecules polar ordered in the N$_F$ phase, as discussed below and in *SI Appendix* **Secs. S9,S10**. This $P$ values is roughly six times larger than the highest polarization ever achieved in tilted calamitic or bent-core chiral smectic LCs [42,43]; is comparable to that found in polar columnar phases [44]; and is well within the range exhibited by solid-state oxide [45] and organic [46] ferroelectrics. This result, combined with our textural observations, indicate that the $N_F$ phase is a 3D fluid, macroscopically homogeneous, polar uniaxial nematic phase. The agreement with the spontaneous $P$ measured from the polar twist threshold indicates that this is the ferroelectric state.

Given these very large polarization values, we summarize here and detail in *SI Appendix* **Sec. S2** several of the relevant features of high-polarization electro-optic, electrostatic, and elastic behaviors developed in the study of chiral smectic ferroelectric LCs, which can now be expected for the $N_F$, some of which are reported here. These include: (*i*) <u>Polar Twist Freedericksz transition</u> - (**Figs. 1,S5**). (*ii*) <u>Boundary penetration</u> [47] - The polar coupling to field limits the distance in which boundary- or defect-preferred orientations transition into bulk field-preferred orientations. This "penetration" length $\xi_E = \sqrt{K/PE} \sim 1$ μm for P = 6 μC/cm$^2$ and an applied field $E$ = 1 V/cm (**Fig. 4**). (*iii*) <u>Block polarization reorientation and expulsion of splay (splay-elastic stiffening)</u> [48,49,50,51,52,53,54,55] - A second "self penetration" length, $\xi_P = \sqrt{\varepsilon K/P^2} \sim 0.1$ nm is the scale above which the polarization field can spontaneously expel spatial variation of orientation ($\nabla \cdot \mathbf{P}(r) = \rho_P$) that produces space charge, $\rho_P$. The result is that low-energy elastic distortions of the *n,P* couple are bend or twist only, with splay of *n(r)* and *P(r)* expelled from the bulk and confined to narrow reorientation walls, as in **Figs. 5A,B** and **S4**. The resulting polarization "blocks" can effectively screen applied field, producing, for example, the large threshold field required for the splay-bend Freedericksz transition in the $N_F$ phase (*SI Appendix* **Sec. S6**). (*iv*) <u>Field-induced torques proportional to E</u> - The balance of field-induced torque proportional to *PE* with viscous torques gives a characteristic reorientation risetime on the order of $\tau = \gamma_1/PE$, where $\gamma_1$ is the nematic rotational viscosity [41,56]. The risetime $\tau = \gamma_1/PE$ is ~ 0.1$\Delta t$, where $\Delta t$ is the reversal time (**Fig. 3C**), giving a value of $\gamma_1 \sim$ 0.1 Pa sec, comparable to that of 5CB at *T* =25ºC. (*v*) <u>The N–$N_F$ phase transition</u> is strongly affected by the polarization self-interaction, which suppresses the longitudinal modulation $\delta P_z$ of **P** [50], producing the strong anisotropy of the polarization fluctuations in the N phase, and rendering the transition mean-field (*SI Appendix* **Sec. S3**).

The electro-optic response of the $N_F$ phase in a cell an electric field applied in-plane shows uniquely polar features, with *P(r)* reorienting in the (*y,z*) plane through an azimuthal angle *φ(r)* about *x* determined by the local surface, elastic and electric torques. Buffed surfaces stabilize two planar-aligned states (*φ*=0 and *φ*=π) with opposite signs of *P(r)*, so the cell has four stable states, two that are uniform and two that are twisted, illustrated in **Fig. 4**. These equilibrium states are separated by π surface disclination walls (magenta dots in the section drawing of **Fig. 4**). If complete polarization reversal is to be achieved by an applied field, **P** must be switched on both surfaces. We refer to domain boundaries such as these, where both *n* and *P* reorient but maintain a fixed relative sign, as Polarization-director (*P-n*) Disclinations (***Pn*Ds**). If the vectorial representation *n(r)* is used to describe the nematic texture, then with only *Pn*Ds the local orientation of *P(r)* relative to *n(r)* will be either parallel or antiparallel everywhere. In **Fig. 4**, an applied field preferring the uniform (U) dark state is deforming a central (magenta) domain that had polarization that was initially opposed to the field and is now partially rotated toward it (green vectors). This central domain forms a twisted-untwisted (TU) state in which the director twists along *x* from the surface-preferred alignment parallel to *z* at one cell plate through azimuthal angle *φ(x)* to the field-aligned orientation in the mid-plane of the cell, and then twists back to the surface-preferred alignment on the other glass plate. The field causes the central domain to shrink, moving and eventually eliminating the disclination walls in order to achieve complete polarization reversal. The motion of the walls on the two cell surfaces is different because the pinning strengths are different and spatially inhomogeneous, with the result that they do not remain in register, leading to the formation of the (olive) left- and (gold) right-handed twisted states $T_L$ and $T_R$ seen surrounding the central domain in *Figs. 4A,B*. The color of the central TU domain (green, blue, or pink in *Fig. 4C*) varies with the degree of field-induced reorientation of *n,P* in the sample mid-plane.

Several other modes of field-induced polarization reversal are shown in *Fig. 5* and *SI Appendix Fig. S4*. The initial response of a uniformly aligned region to an increasing in-plane DC field in the range $0 < E < 2$ V/cm that opposes the local polarization is to form a zig-zag modulation in the orientation of *n(r)* and *P(r)*, illustrated in *Fig. S4*, in which the non-zero spatial variation is $\partial n(r)/\partial z$, along the director, making it a bend wave. Bend has a lower polarization space charge energy cost than a splay wave (which, with non-zero $\partial n(r)/\partial x$, would generate stripes parallel to *n* rather than normal to it). As the field strength is increased, the degree of reorientation increases and distinct boundaries appear between the half-periods of the modulation, separating stripes of uniform internal orientation. Fields of a few V/cm drive complete (+π, -π, +π, -π) reorientation of *n* in the cell mid-plane, at which point these boundaries become 2π walls sub-optical in resolution, ~$ξ_P$ in width. This process is even more dramatic with dynamic driving, as illustrated in *Fig. 5A*, which shows snapshots during field reversals generated by a 5 Hz AC triangle wave voltage with different amplitudes. As the field amplitude increases from 0 to $E_p = 10$ V/cm, the stripes become very regular and narrowly spaced. The herringbone pattern of polarization in the stripes gives an overall structure where $P_z$ is constant, ensuring that there is no net polarization charge at the stripe boundaries, and where the backflow induced in each stripe matches that of its neighbor. The field is not strong enough to reverse the surfaces in this case. A different reorientation mode is illustrated in *Fig. 5B*, where field reversal leads to the formation of polygonal domains in which charge-stabilized areas of uniform *P* are bounded by sharp domain boundaries, each oriented along a vector *l* such that *P·l* has the same value on either side of the boundary, as shown in panel 4. As with the stripes above, this geometry reduces the net polarization charge on the line. Similar structures are found in high-*P* chiral smectic ferroelectrics [57] and in ferromagnets [58]. The textures of the charge-stabilized domains can also be employed to visualize directly the reorientation of *P(r)* under applied field, as shown in *Fig. 5C*, where a circular air bubble enables tracking of the polar orientation of *n(r)* during field reversal. The director is anchored tangent to the bubble surface, resulting in a director field that is largely bent around the bubble, with splay concentrated in two 180° wedge disclinations (red dots). *n(r)* in the area surrounding the bubble is parallel to the line connecting these defects. The *n(r),P(r)* structures observed with increasing reversal field strength are sketched below each panel.

*Nematic Ferroelectrohydrodynamics* – The polarization density of the $N_F$ phase creates a fluid which is extraordinarily responsive both to external applied fields and to its internally generated polarization space charge. While the discussion above has focused on the effects of field-induced molecular reorientation, the most interesting and useful effects of the $N_F$ may be its ferroelectrohydrodynamic or ferroelectrorheological behavior, exemplified by the observations shown in *Fig. 6* and *SI Appendix Sec. S7*. In this experiment, RM734 is filled into a $t = 10$ μm cell with random-planar alignment of *n*. An in-plane electric field is applied using a pair of gold electrodes evaporated with a $d = 60$ μm gap onto one glass plate, visible at the bottom of *Fig. 6A*. A square wave voltage with $V_P = 5$ V applied between the electrodes generates an electric field distribution where $E(r)$ is uniform in the electrode gap and in the surrounding area is directed along half-circular arcs centered on the gap (*Fig. S15A* inset). In the $N_F$ phase, this field induces flow of localized defects (*Fig. 6B*) and their surrounding fluid with a velocity field $v(r,t)$ locally parallel to $E(r)$ and changing direction with the field, suggesting an electric body force density $F(r) = \rho(r)E(r)$, where $\rho(r)$ is a positive electric charge density. When $E(r)$ goes through $E = 0$ during field reversal, flow ceases and the director field breaks up into *P*-reversal bend-domain bands like those shown in *Fig. 5A* as it rotates alternately through $+\pi$ and $-\pi$, giving the radial texture seen in *Fig. 6A*. Thus, dynamically $P(r,t)$ is everywhere parallel to $E(r,t)$ and $v(r,t)$ when voltage is applied. The fact that the product $P(r) \cdot E(r)$ is unchanged by applied field reversal and yet $v(r)$ changes sign, indicates that $\rho(r)$ does not change sign with $P(r)$, i.e., that the driving has caused the fluid to become charged. The structure of polarization reversal bands in the neighborhood of the electrode gap is shown in detail in *SI Appendix Fig. S16*.

We measured $v_P$, the initial value of the defect velocity upon field reversal at the location indicated in *Fig. 6A*. This velocity depends dramatically on temperature, as shown in *Fig. 6C*, with flow being essentially absent in the N phase and commencing upon cooling through the N–$N_F$ transition. The velocity eventually decreases with decreasing $T$, presumably because of the increasing viscosity of the LC.

The experiments show that applied electric field promotes the creation of regions with positive charge density. Charging of the $N_F$ by AC applied fields is to be expected due to the bulk polarity of the phase. Electrode surfaces contact $N_F$ material where the direction of *P* alternates in time. The $N_F$, because of its polar symmetry, has diode-like, polarity-dependent resistance that can also depend on the sign and nature of the charge carrier. The bulk charge mobility along $z$ in the $N_F$ phase may also depend on field direction. Beyond this there will be a variety of charging effects due to the linear coupling of *P* and flow. Let us consider, for example, steady, incompressible nematic laminar flow, then the director is generally nearly parallel to the velocity and $v(r) = v(r)n(r)$. Since $\nabla \cdot v(r) = 0$ we have $\nabla \cdot n(s) = [\ln v(s)]/\partial s$, where $s$ is the position variable along the flow: where the velocity increases the director splays inward. But in the $N_F$ phase we have $P(r) = Pn(r)$, where $P$ is the constant polarization magnitude, so that laminar flow produces polarization charge density $\rho_P(s) = P\nabla \cdot n(s) = P\partial[\ln v(s)]/\partial s$, the sign of which depends on whether $P$ is aligned along $v$ or opposed to it. Complex flows will thus produce complex patterns of polarization charge. Reorientation of *P* generates displacement current, $J = \partial P/\partial t$, which is locally normal to $P(r)$ and, if driven by electric field, gives a highly anisotropic contribution to the net electrical conductivity, $\sigma_\perp = P^2/\gamma_1$ for $E \perp P$, and $\sigma_\parallel = 0$ for $E \parallel P$ [43]. For RM734, we obtain $\sigma_\perp \sim 10^{-3}/\Omega$cm, which is in the semiconducting range. Under these circumstances, accumulation of one sign of charge in the fluid can occur when an applied AC field gets out of phase with polarization reversal. Additional inherent asymmetries, such as differences in mobility or chemical character between positive and negative ionic impurities, or an intrinsic tendency for splay distortion of the $P(r)$ field itself, can also contribute.

*Atomistic Molecular Dynamics (MD) Simulation*

We carried out MD simulations directed toward gaining an understanding of how features of molecular architecture, interactions, and correlations are related to the polar ordering of the $N_F$ phase. These calculations used a simulation box containing 384 RM734 molecules (*SI Appendix* **Fig. S18**) with periodic boundary conditions, equilibrated in the *NPT* ensemble at $p = 1$ atm for a range of temperatures spanning the N and $N_F$ phases, using the APPLE&P force field [59] successfully applied in previous studies of nematic [60] and twist-bend [61] phases. More details of the simulations can be found in *SI Appendix* **Secs. S9,S10**.

The simulations probe the equilibration of RM734 in two distinct condensed LC states: (*i*) ***POLAR*** (*POL*) – a polar nematic state generated by equilibrating a starting condition that is perfectly ordered in the +z direction of the polar molecular long-axis vectors *u*, which point from the nitro (O) to the methyl (H) end of each molecule as seen in *Fig. 7A*; (*ii*) ***NONPOLAR*** (*NONPOL*) – a weakly polar, nematic state generated by equilibrating a starting condition with no net polar order (50%/50% division of the *u* vectors along +z/-z). Each molecule can then be labeled as OH or HO, depending on its orientation (whether *u* is along +z or -z, respectively). Equilibration with respect to the internal molecular arrangements of these two systems is readily achieved through orientational and internal molecular fluctuations as well as diffusive molecular motion. Their equilibrated states are distinct, however: the steric packing of the anisotropic molecules makes head-to-tail molecular flipping events extremely rare during the simulation space-time volumes, so the equilibrations obtained have an effective constraint of no molecular flipping.

These *POL* and *NONPOL* states represent the extremes of equilibrated polar order and polar disorder in the simulation volume. Limiting the simulations to these states, i.e., not considering molecular flips in a simulation of polar order, may seem like a significant shortcoming. However, when we use the *POL* state to calculate *P*, we obtain polarization densities that match those of RM734 at low *T* (see ***Fig. 3B***), implying that at low temperatures the ordering of RM734 becomes that of the simulated *POL* state, making this an ideal model system for exploration of the molecular correlations leading to polar order.

The positional pair-correlation functions, $g_P(\rho,z)$ and $g_{NP}(\rho,z)$ of the equilibrated *POL* and *NONPOL* systems are shown in ***Figures 7B-E***, where for the *NONPOL* system $g_{NP}(\rho,z) = g_{NP}^{par}(\rho,z) + g_{NP}^{anti}(\rho,z)$ is the sum of the correlations between the molecular pairs with relative parallel or antiparallel orientations of *u*. The $g(\rho,z)$ are $\varphi$-averaged conditional probability densities of molecular centers around a molecule with its center at the origin and long axis along z and thus are uniaxially symmetric in $(\rho,\varphi,z)$ cylindrical coordinates, reflecting the uniaxial symmetry of the N and $N_F$ phases. They all exhibit a molecule-shaped, low-density region ($g(\rho,z) \sim 0$) around the origin resulting from the steric overlap exclusion of the molecules; an asymptotic constant density at large $\rho$; and distinct peaks indicating preferred modes of molecular packing. The normalized average density is $<g(\rho,z)> = 1$. If we were to consider a similarly equilibrated system of rods marked "O" and "H" on their ends but which were otherwise symmetric (e.g., hard spherocylinders marked as either OH or HO) then both of the orientational states would have identical pair-correlation functions, $g_P(\rho,z) = g_{NP}(\rho,z) = 2g_{NP}^{par}(\rho,z) = 2g_{NP}^{anti}(\rho,z)$. The $g(\rho,z)$ of RM734, in contrast, show a number of striking differences that directly exhibit the effects of its structural and electrostatic polarity on the packing of neighboring molecules.

The *POL* $g_P(\rho,z)$ in *Fig. 7B* shows the equilibrated, local molecular packing preference in the limit of polar order, i.e., in the system with the maximum number of contacts between like-oriented molecules. Its prominent features are sharp, on-axis arcs at ($\rho = 0$, $z = \pm 22$ Å), indicating on-average coaxial molecular association into polar chain-like (OH-OH-OH) associations having a center-to-center spacing along $z$ of the molecular length, 22 Å, stabilized by the electrostatic attraction of the nitro and methoxy ends of the molecules (*Fig. 7F*); and off-axis peaks at $\rho = 5$ Å, $z = \pm 6$ Å, indicating polar side-by side association (*Fig. 7G*). These correlations indicate that specific electrostatic interactions between oppositely charged groups on the two molecules (e.g., between positively charged terminal or lateral methoxy H atoms and negatively charged nitro O atoms) play a dominant role in stabilizing such pair configurations.

The *NONPOL* system enforces the maximum number of molecular contacts between molecules of opposite orientation. In this situation of maximum polar disorder, possible equilibrated molecular correlations could range from being (*i*) dominantly antiparallel end-to-end (e.g., OH-HO-OH chains, with side-to-side polar correlations, as in the bilayer smectics of strongly polar molecules [62]); to being (*ii*) polar end-to-end (a mixture of OH-OH-OH and HO-HO-HO chains with the OH-HO interactions side-by-side). RM734 is distinctly in the latter category, as, remarkably, the principal polar ordering motifs of *Figs. 7F,G* are even stronger in the *NONPOL* system than in the *POL* (compare *Figs. 7B* and *7D*), and the antipolar correlations are largely side-by side. The OH-HO end-to-end antipolar association depicted in *Fig 7J* is present but weak, as is the HO-OH end-to-end pairing of *Fig. 7K*. The latter is dominant in the crystal phase [32] but not as a mode of achieving antipolar ordering in the *NONPOL* system. It appears from these results that the polar correlations identified in the *POL* system, and persisting in the *NONPOL* system in the maximal presence of enforced polar disorder, could be solely responsible for the stabilizing the $N_F$ phase. Prominent features of the correlations are sharp, on-axis arcs at ($\rho = 0$, $z = \pm 22$ Å), indicating on-average coaxial molecular association into polar chain-like (OH-OH-OH) associations, also with a center-to-center spacing along $z$ of the molecular length, 22 Å, and stabilized by the electrostatic attraction of the nitro and methoxy ends of the molecules (*Fig. 7F*); and off-axis peaks at $\rho = 5$ Å, $z = \pm 6$ Å, indicating polar side-by side association (*Fig. 7G*). These correlations indicate that specific electrostatic interactions between oppositely charged groups distributed throughout the two molecules (e.g., between positively charged terminal or lateral methoxy H atoms and negatively charged nitro O atoms) play the dominant role in stabilizing polar pair configurations.

The *POL* simulation equilibrates a state in which end-to-end flipping is kinetically arrested and the periodic boundary conditions suppress long-wavelength orientation fluctuations ($\lambda_x > 55$ Å and $\lambda_z > 70$ Å). The remnant short-ranged fluctuations lead to pair correlations which, as we have shown in ***Fig. 7***, are confined to the volume $\rho < 10$ Å and $z < 30$ Å about the origin, molecular neighbor separation scales which are well within the dimensions of the simulation box. These conditions create a *"plupolar" (plus quam pola*r [63]) equilibrium state of constrained polar ordering yielding the simulated $P$ values shown in ***Fig. 3*** (open circles). Comparing these polarization values with the RM734 data shows (i) that in the *plupolar* state, the fluctuations that lead to the phase transition are clearly suppressed, while the remnant short range fluctuations give a $P$ value exhibiting only a weak dependence on temperature; and (ii) that this $P$ gives a good account of the polarization density of the $N_F$ at low temperature, evidence that at low $T$ the $N_F$ phase approaches a comparable *plupolar*-like condition where there are only short-range fluctuations, and where the simulated $g(\rho,z)$ faithfully represent their correlations.

## *Materials and Methods*

***Synthesis of RM734*** – 4-[(4-nitrophenoxy)carbonyl]phenyl 2,4-dimethoxybenzoate (RM734, ***Fig. 1A***) is a rod-shaped mesogen first synthesized by Mandle et al. [31]. It was reported to have an isotropic (I) phase and two additional phases with nematic character, with transition temperatures as follows:
I – 187ºC – N – 133ºC – $N_X$ – X. Our preparation is based on general synthetic reactions and uses procedures only slightly modified from those described in the literature cited (*SI Appendix **Sec. S1***).

***Observations of Response to Applied Electric Field*** - Experimental cells were made by filling LC samples between glass plates coated with lithographically patterned ITO or gold electrodes and spaced to a desired gap, *t*. Both transparent capacitor and in-plane electrode geometries were employed. Experiments were performed in temperature controlled environments, with electro-optic observations carried out using depolarized transmission light microscopy with cells mounted on the rotary stage of a research microscope and imaged in transmitted light between polarizers. The sign and magnitude of the in-plane birefringence were determined using a Berek compensator (***Fig. S17***). Polarization measurements were made by using transimpedance electronics to integrate the current in response to an applied electric field, using several in-plane geometries and a glass capillary cell with coaxial electrodes (*SI Appendix **Sec. S1***).

## *Acknowledgements*

This work was supported Materials Research Science and Engineering Center (MRSEC) Grant DMR 1420736, and by NSF Condensed Matter Physics Grant DMR 1710711. Leo Radzihovsky was supported by a Simons Investigator Award from the Simons Foundation. The authors also acknowledge the Center of High Performance Computing at the University of Utah for allocation of computing resources.

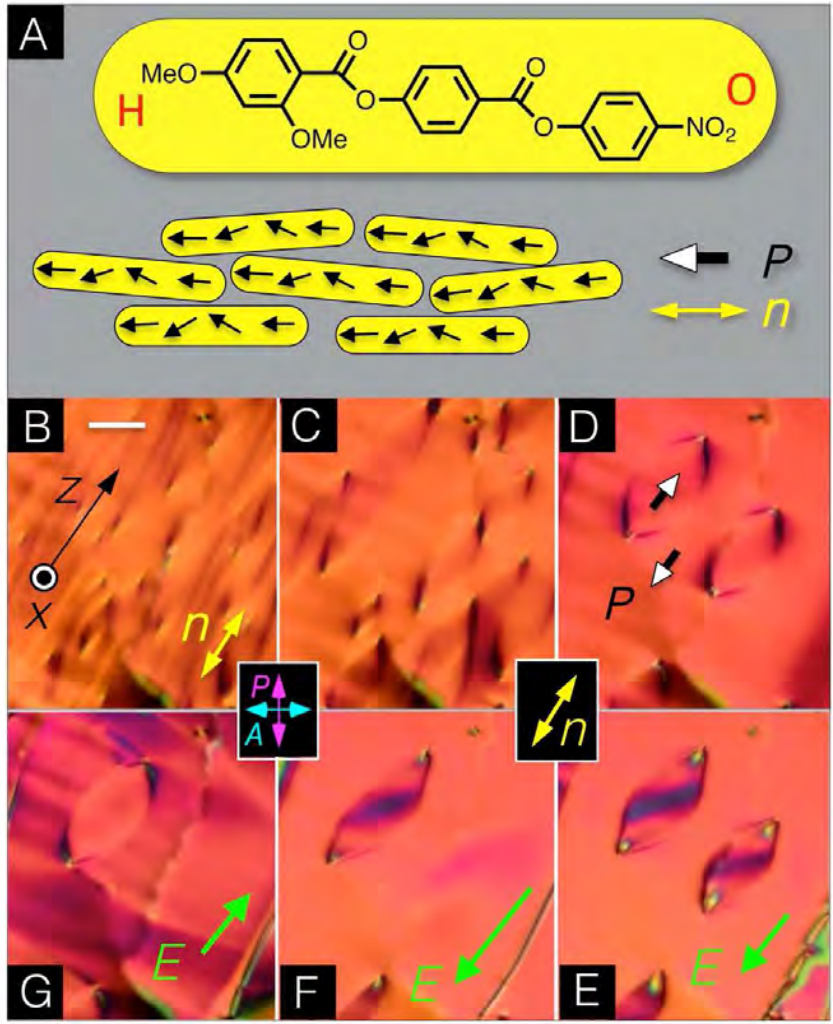

*Figure 1*: Ferroelectric nematic phase. (**A**) Structure of RM734 and schematic of molecular alignment in the ferroelectric nematic ($N_F$) phase. The molecular organization is translationally symmetric in 3D and macroscopically uniaxial, with local mean molecular long axis, *n(r)*, aligned generally along the buffing direction *z*; and polar, with a local, mean molecular dipole orientation, *P(r)* along *n*. H and O are used to represent the methoxy and nitro- ends of the molecule respectively. (**B–G**) DTLM images showing electro-optic evidence for ferroelectricity in a planar-aligned cell of RM734 in the $N_F$ phase (*t* = 11 μm thick). In the higher temperature N phase, *P(r)* = 0 but when cooled into the $N_F$ phase without an applied field, RM734 spontaneously forms macroscopic domains with *P* > 0 or *P* < 0. When slowly cooled below the $N_F$ phase transition at *T* = 133°C, the initial texture (**B**) coarsens into a pattern of domains with distinct boundaries (**C**). (**D–G**) *T* = 120°C. Starting from (**D**) with no field, application of an ultra-small in-plane test field |$E_z$| ~ 0.5 V/cm along the buffing direction produces reversible reorientation of *P* without changing its magnitude. (**E,F**) Application of a negative $E_z$ starts the in-plane reorientation of *n(r)* about *x* inside the domains, producing the dark bands there, while (**G**) positive $E_z$ produces reorientation outside of the domains, proving that these regions are of opposite polarization. The *E* ~ 1 V/cm threshold field for this reorientation indicates that *n(r)* in these domains is coupled to *E* by a polarization *P* ~ 5 μC/cm$^2$, which is comparable to the bulk polarization density measured electronically. The higher applied field in (**F**) has moved the boundary of one lenticular domain to increase the area with the field-preferred orientation, effecting a hysteretic reversal of *P(r)*. Scale bar = 100 μm.

*14*

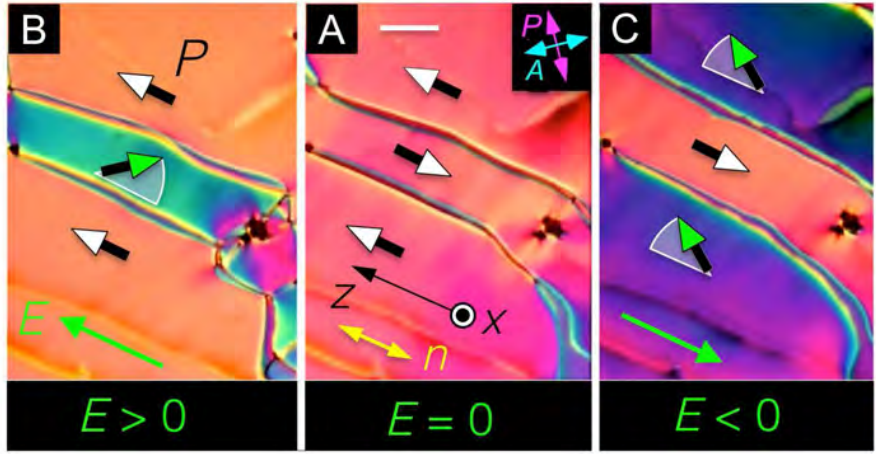

*Figure 2*: DTLM images showing polar Freedericksz twist transition in ferroelectric domains with opposite polar orientation at $T = 120°C$. These domains, grown field-free upon cooling from the N phase to this temperature, have a polarization density ***P***. (**A**) Field-free initial state showing three domains separated by domain walls, each domain having ***n***(***r***) along the buffing direction ***z***.   (**B**) Application of an ultra-small, positive test field $E_z = 1$ V/cm induces a birefringence color change resulting from in-plane reorientation of ***n***(***r***) in the center domain, leaving the upper and lower domains unchanged. (**C**) Application of $E_z < 0$ induces an in-plane reorientation of ***n***(***r***) in the upper and lower domains. There is little optical change or reorientation in the central domain. If the field is returned to $E = 0$, the system returns to the starting state (**A**). These observations demonstrate that the domains are polar and also enable the absolute determination of the direction of ***P***(***r***): domains that have the orientation preferred by the applied field do not reorient.  In this experiment, ***P***(***r***) and ***n***(***r***) within the domains rotate about ***x*** but the field is not large enough to move the domain walls, which are pinned by the surfaces.  The polarization vectors (shaded green) and circular arcs (white) depict the field-induced reorientation of ***P***(***r***) in the mid-plane of the cell: ***P***(***r***) does not reorient at the surfaces in this experiment, remaining parallel to the buffing direction.  These field-induced reorientations with ***P***(***r***) starting nearly antiparallel to ***E*** are polar azimuthal Freedericksz transitions. The threshold field, $E_P = (\pi/t)^2(K_T/P)$, estimated using the measured $P \sim 5$ $\mu C/cm^2$ at $T = 120°C$ (see *Fig. 3*), is $E_P \sim 1$ V/cm, comparable to the fields employed here.  $t = 11$ $\mu$m. Scale bar = 100 $\mu$m.

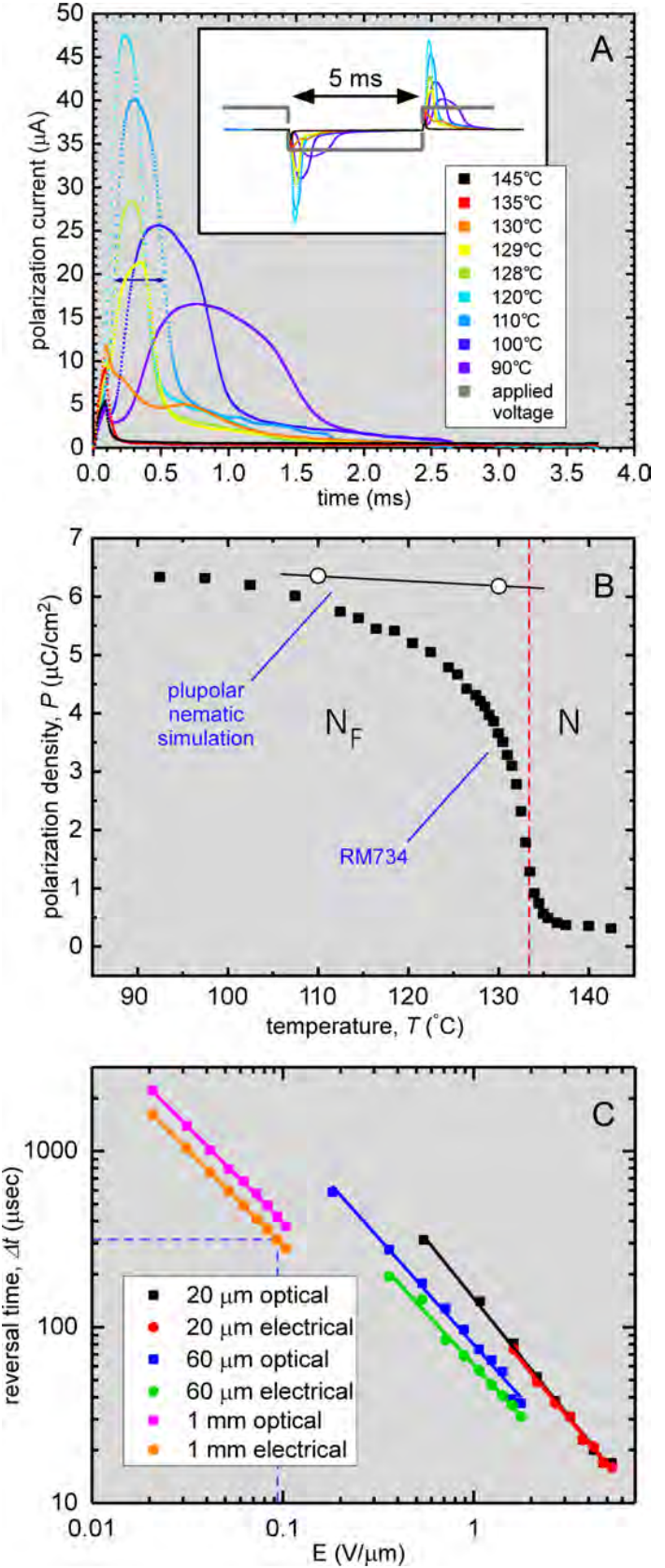

*Figure 3*: Characteristics of polarization reversal in an applied field. (**A**) Temperature dependence of the cell current with a 200 Hz square wave generating a field $E_p = 95$ V/mm applied in-plane to a $t = 15$ µm-thick cell with 1 cm-wide ITO electrodes spaced by $d = 1$ mm. In the I and N phases ($T \geq 133°C$), the current is small and capacitive. On cooling into the $N_F$ phase, an additional current contribution appears, the area of which is independent of voltage and is equal to the net polarization reversal charge, $Q = 2PA$, where $A = 15$ µm x 1 cm is the effective cross-sectional area of the volume of LC material reoriented by the applied field. In the $N_F$ phase, the polarization reversal current becomes a distinct peak that grows in area on cooling, indicative of an increasing polarization density, and the reorientation takes place more slowly, reflecting the increase of orientational viscosity. The double-headed arrow shows the reversal time at $T = 110°C$ [dashed drop lines in (**C**)]. (**B**) The polarization density $P$ of RM734 measured on cooling (black squares) saturates at $P \sim 6$ µC/cm² at the lowest temperatures. The open circles are values of $P$ of the *plupolar* nematic calculated from the *POL* MD simulation of the $N_F$ phase (*Fig. 7*, *SI Appendix Sec. S9,10*). In the *plupolar* nematic, long-wavelength orientation fluctuations are suppressed, giving a $P$ value determined by molecular-scale fluctuations and local packing. RM734 approaches the *plupolar* condition at low $T$. The region near the transition has not been studied in detail. (**C**) Field dependence of the reversal time $\Delta t$, taken as the full width at half-height of the polarization or optical reversal pulse following a field step in a 100 Hz, bipolar, square-pulse train of peak amplitude $E_p$ in planar-aligned cells with in-plane electrodes spaced by $d = 20$ µm, 60 µm, and 1 mm at $T = 110°C$. The reversal time scales as $1/E_p$ as expected for reorientation driven by ferroelectric torques. The dashed lines identify the measurement with $E_p = 95$ V/mm highlighted in (**A**). The risetime $\tau = \gamma_1/PE$ is $\sim 0.1\Delta t$, giving a value of $\gamma_1 \sim 0.1$ Pa-s, comparable to the viscosity of 5CB at $T = 25°C$.

*16*

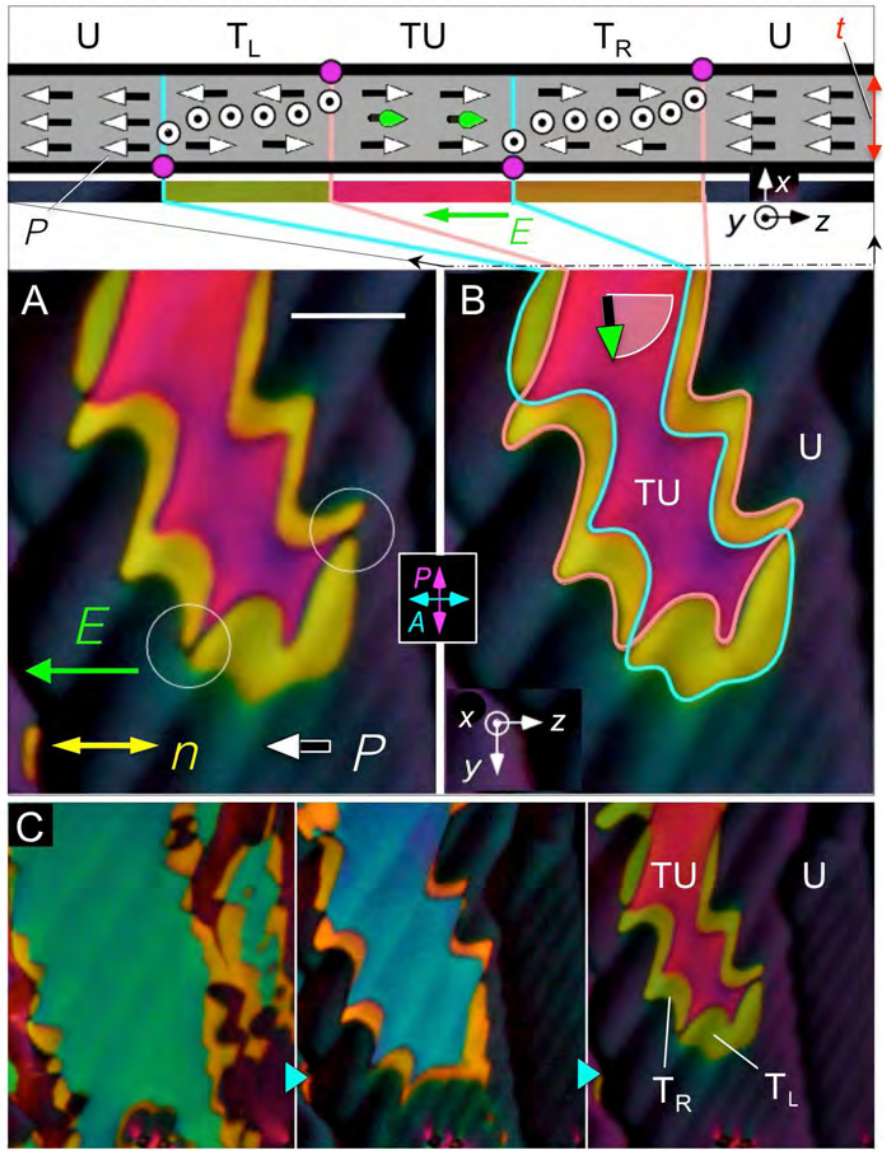

*Figure 4*: DTLM images of a large, twisted domain with surface polarization pointing to the right, surrounded by a uniform region with surface and bulk polarization pointing to the left, in the direction of an applied field. (**A**) Twisted domain (magenta) and structural elements *P*, *n*, and *E*. (**B**) The section drawing shows the 2D structure of the cell in the *x,z* plane along the top edge of the image: the uniform (U), field-preferred state of the background, the surface orientations reversing at the boundaries of the central domain, *P* in the twisted-untwisted (TU) state in the center of the domain, with the orientation in the middle of the cell indicated by green vectors, and the intermediate left- and right-handed twisted states T$_L$ and T$_R$ (olive and gold). π surface disclination lines (magenta dots) mediate polarization reorientation at the top (pink line) and bottom (cyan line) cell plates. Where the surface disclination lines overlap, the director is uniformly oriented along *y* through the thickness of the cell, giving extinction between the crossed polarizers (dark spots circled in (**A**)). In the absence of applied field, the left and right surface polarization states are energetically equivalent. (**C**) The central domain shrinks with increasing *E* field. The birefringence color changes from green to blue to pink as the rotation of *P* in the middle of the cell increases. $T = 120°C$. $t = 11$ μm. Scale bar = 50 μm.

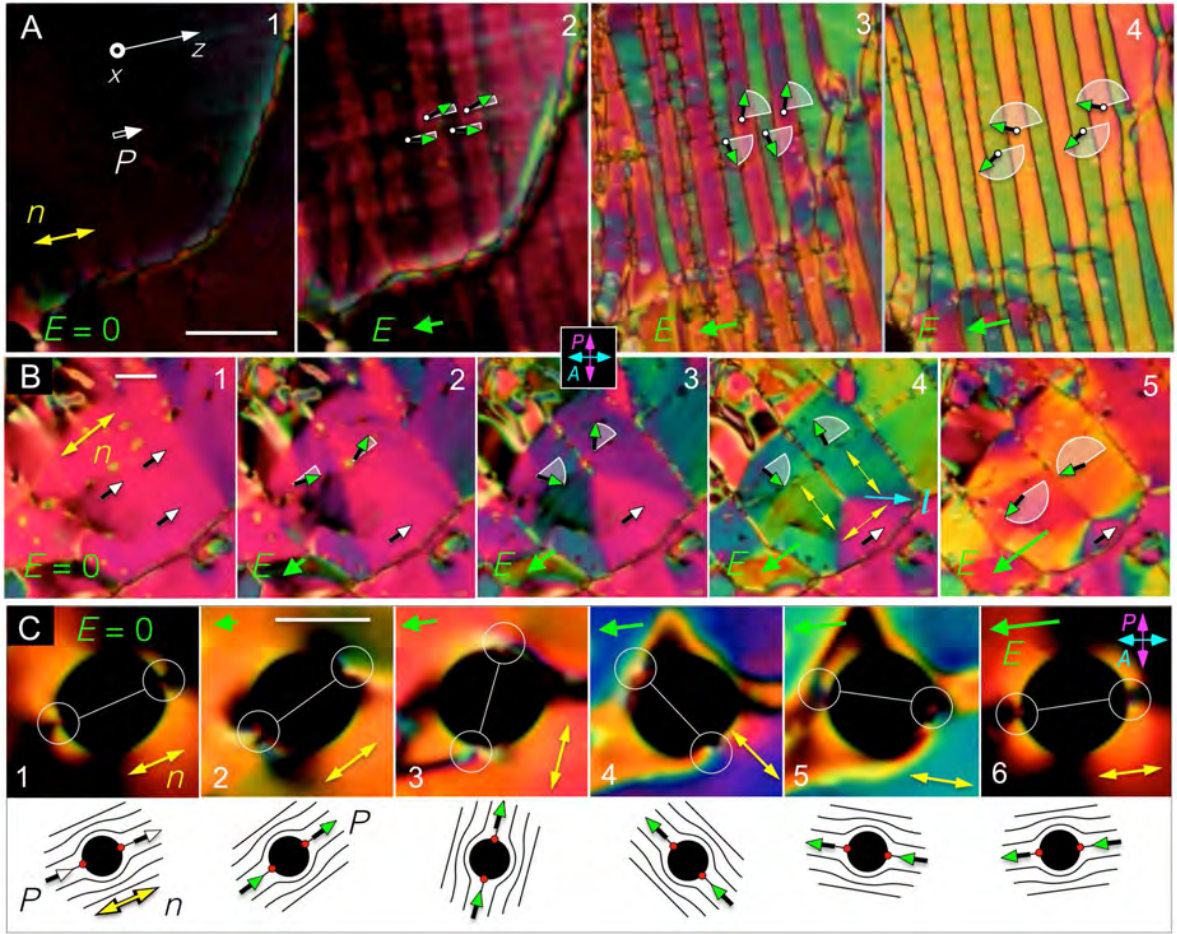

*Figure 5*: Common polarization reversal scenarios in RM734. Field-induced reorientation of **P** is indicated schematically using white arcs and green vectors. (**A**) Stripe formation. Applying a 5 Hz triangle-wave electric field with peak amplitudes in the range $0 < E_p < 10$ V/cm to a region with an initially uniform in-plane director (panel 1) induces a periodic modulation in the orientation of *n*(*r*) and *P*(*r*) along *z* (a director bend wave, panel 2) whenever the field changes sign. As the applied field strength is increased (panels 3,4), the stripes form with sharper boundaries and have uniform internal orientation determined by the field strength. The zig-zag arrangement of the director in successive stripes ensures that the normal component of *P* is constant across the stripe boundaries, so that there is no net polarization charge there. (**B**) Polygonal domains. During field reversal, polarization charge effects alternatively lead to the formation of tile-like domains with uniform *n*(*r*). These polygons have sharp domain boundaries that are oriented such that *P*·*l*, where *l* is along the boundary, is the same on both sides of the boundary, reducing space charge. The angular jump in *n*(*r*) across the boundary highlighted in panel 4 is 90°. (**C**) Director field reorientation around inclusions. Air bubbles in the cell can be used to track the orientation of *n*(*r*) in a reversing field. The director field near the bubble, sketched below each panel, is locally distorted, bending around the inclusion with splay deformations confined to two 180° wedge disclinations (red dots) located at opposite ends of the bubble. The blue color in panels 4 and 5 is indicative of a TU state of the kind shown in *Fig. 4*, with a surface disclination then moving out from the bubble boundary to give the final, uniform state seen in panel 6. Scale bars: A = 100 μm; B = 100 μm; C = 50 μm.

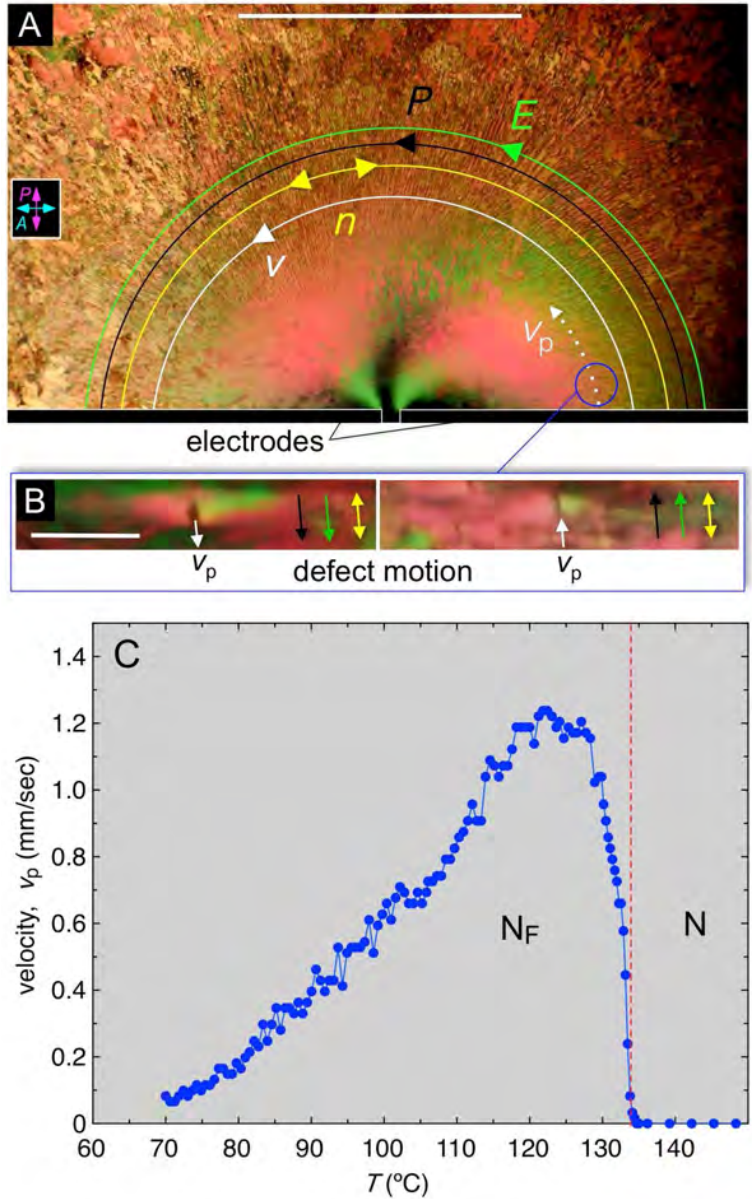

*Figure 6*: Field-induced flow in the ferroelectric nematic phase. (**A**) DTLM image of a $t = 10$ μm-thick, planar-aligned cell of RM734 between untreated glass plates, in the $N_F$ phase at $T = 120$°C. The black bars at the bottom are two evaporated gold electrodes on one of the plates, separated by a $d = 60$ μm gap. The electrodes are outlined in white for clarity. Only the upper edges of the electrodes and the adjacent part of the cell are shown. A square-wave voltage with $V_p = 3$ V, 0.1 Hz is applied to the electrodes, producing an electric field in the plane of the cell. This field drives a pattern of defect motion and fluid flow over the entire field of view, with the defect velocity $v(r)$ (white arrows) parallel to the applied field, $E(r)$ (green), which is tangent to half-circles centered on the electrode gap (see *SI Appendix Fig. S12*). Where the defects are dense, their motion transports the surrounding fluid. When the field is applied, the entire region shown here moves along the field lines. This image, captured during field reversal, shows a periodic array of bend domain walls normal to the director (yellow) and the applied field, as in *Fig. 5A*, in this case along radial lines. (**B**) Typical defect in the texture moving along the applied field direction (down in the first image, up in the second), in the location circled in (**A**). (**C**) Temperature dependence of the magnitude of the initial defect velocity along the white dashed track in (**A**) following a field reversal. There is no flow in the N phase but on cooling into the $N_F$ phase, the velocity increases rapidly with increasing $P$ before decreasing again at lower $T$ because of the increasing viscosity. A similar dependence on $T$ is observed whether heating or cooling. Scale bars: A = 1 mm, B = 100 μm.

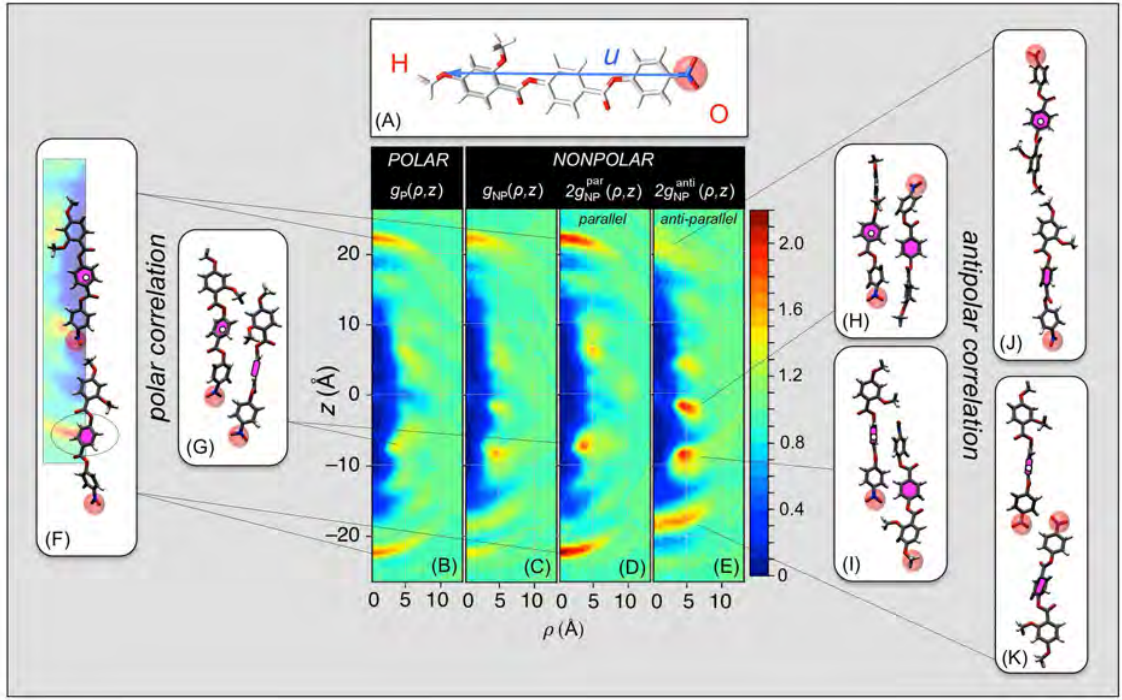

*Figure 7*: Molecular pair correlations revealed by atomistic molecular dynamic simulations designed specifically to explore the molecular interactions and resulting positional/orientational correlations responsible for the polar molecular ordering of RM734, shown in (**A**). A nanoscale volume containing 384 molecules is equilibrated in these simulations into two distinct LC states: a ***POLAR*** system with all polar molecular long axes, ***u***, along +z, and a ***NONPOLAR*** system with half along +z and half along -z. Equilibration of the molecular conformation and packing is readily achieved but end-to-end flips are rare, so the simulated states remain in the polar or non-polar limit of equilibrated nematic order, respectively. (**B,F,G**) The *POL* simulation shows directly the dominant pair correlations adopted by molecules that are polar ordered, in the form of conditional probability densities, $g(\rho,z)$, of molecular centers (magenta fill) around a molecule with its center (white dots) at the origin and long axis ***u*** along z. The $g(\rho,z)$ are $\varphi$-averaged to be uniaxially symmetric, reflecting the uniaxial symmetry of the N and N$_F$ phases. They exhibit a molecule-shaped, low-density region ($g(\rho,z) \sim 0$) around the origin resulting from the steric overlap exclusion of the molecules; an asymptotic constant value at large $\rho$ giving the normalized average density ($g(\rho,z) = 1$); and distinct peaks indicating preferred modes of molecular packing. This analysis reveals two principal preferred packing modes in the *POL* system: (**B,F**) polar head-to-tail association stabilized by the attraction of the terminal nitro and methoxy groups, and (**B,G**) polar side-by-side association governed by group charges along the molecule, nitro-lateral methoxy attraction, and steric interactions of the lateral methoxys. (**D,E**) The *NONPOL* system exhibits distinct correlation functions for antiparallel and parallel molecular pairs, $g_{NP}^{anti}(\rho,z)$ and $g_{NP}^{par}(\rho,z)$. (**E,H,I**) The preferred antiparallel packing gives strong side-by-side correlations, governed by group charges along the molecule; and (**E,J,K**) weaker antipolar nitro-nitro end-to-end association. (**D,F,G**) The parallel correlations in the *NONPOL* system are the most relevant to the stability of polar order in the N$_F$ phase, as they are determined by the inherent tendency of the molecular interactions for polar ordering in the presence of enforced polar disorder. Comparison of (**B**) and (**D**) shows identical preferred modes of parallel association in the two systems, with the *POL* system correlations being even stronger in the *NONPOL* system. This is clear evidence that the polar packing motifs giving the correlation functions (**B**) and (**D**), exemplified by the sample *POL* MD configurations (**F**) and (**G**), stabilize the polar order of the ferroelectric nematic phase.

*20*

# SUPPLEMENTARY INFORMATION

*First-Principles Experimental Demonstration of Ferroelectricity in a Thermotropic Nematic Liquid Crystal: Polar Domains and Striking Electro-Optics*

X. Chen[1], E. Korblova[2], D. Dong[3], X. Wei[3], R.F. Shao[1], L. Radzihovsky[1],
M.A. Glaser[1], J.E. Maclennan[1], D. Bedrov[3], D.M. Walba[2], N.A. Clark[1]

[1]*Department of Physics and Soft Materials Research Center,
University of Colorado, Boulder, CO 80309, USA*

[2]*Department of Chemistry and Soft Materials Research Center,
University of Colorado, Boulder, CO 80309, USA*

[3]*Department of Materials Science and Engineering, University of Utah, Salt Lake City, UT 84112, USA
and Soft Materials Research Center,
University of Colorado, Boulder, CO 80309, USA*

*Corresponding Author* - Noel Clark (*noel.clark@colorado.edu*)

*Abstract*

We report the experimental determination of the structure and response to applied electric field of the lower-temperature nematic phase of the calamitic literature compound
4-[(4-nitrophenoxy)carbonyl]phenyl2,4-dimethoxybenzoate (RM734). We exploit its electro-optics to visualize the appearance, in the absence of applied field, of a permanent electric polarization density, manifested as a spontaneously broken symmetry in distinct domains of opposite polar orientation. Polarization reversal is mediated by field-induced domain wall movement, making this phase ferroelectric, a 3D uniaxial nematic having a spontaneous, reorientable, polarization locally parallel to the director. This polarization density saturates at a low temperature value of ~ 6 $\mu C/cm^2$, the largest ever measured for a fluid or glassy material. This polarization is comparable to that of solid state ferroelectrics and is close to the average value obtained by assuming perfect, polar alignment of molecular long axes in the nematic. We find a host of spectacular optical and hydrodynamic effects driven by ultra-low applied field (E~1 V/cm), produced by the coupling of the large polarization to nematic birefringence and flow. Electrostatic self-interaction of the polarization charge renders the transition from the nematic phase mean-field-like and weakly first-order, and controls the director field structure of the ferroelectric phase. Atomistic molecular dynamics simulation reveals short-range polar molecular interactions that favor ferroelectric ordering, including a tendency for head-to-tail association into polar, chain-like assemblies having polar lateral correlations.   These results indicate a significant potential for transformative new nematic science and technology based on the enhanced understanding, development, and exploitation of molecular electrostatic interaction.

*TABLE OF CONTENTS*

***Materials*** – <u>RM734 - 4-[(4-nitrophenoxy)carbonyl]phenyl 2,4-dimethoxybenzoate</u> (***Fig. S1***) is a rod-shaped molecule about 20 Å long and 5 Å in diameter, with a longitudinal electric dipole moment of about 11 Debye. First reported by Mandle et al. [1,2], it was found to melt at $T = 140$°C and have an isotropic (I) phase and two additional phases with nematic-like character, with transition temperatures on cooling as follows: I – 188°C – N – 133°C – $N_X$. Our synthetic scheme, shown in ***Fig. S1***, is based on general synthetic reactions and procedures only slightly modified from those described in the literature cited. The synthesized compound has very similar phase transition temperatures to those reported by Mandle (***Fig. S2***).

Reagents, solvents, and starting materials were used as purchased from qualified suppliers without additional purification. Reactions were performed in oven-dried glassware under an atmosphere of dry argon. Purification by flash chromatography was performed with silica gel (40–63 microns) purchased from Zeochem AG. Analytical thin-layer chromatography (TLC) was performed on silica gel 60 $F_{254}$ TLC plates from Millipore Sigma (Darmstadt, Germany) Compounds were visualized with shortwavelength ultraviolet (UV). Nuclear magnetic resonance (NMR) spectra were obtained using a Bruker Avance-III 300 spectrometer. NMR chemical shifts were referenced to $CHCl_3$ (7.24 ppm for 1H, 77.16 ppm for $^{13}C$).

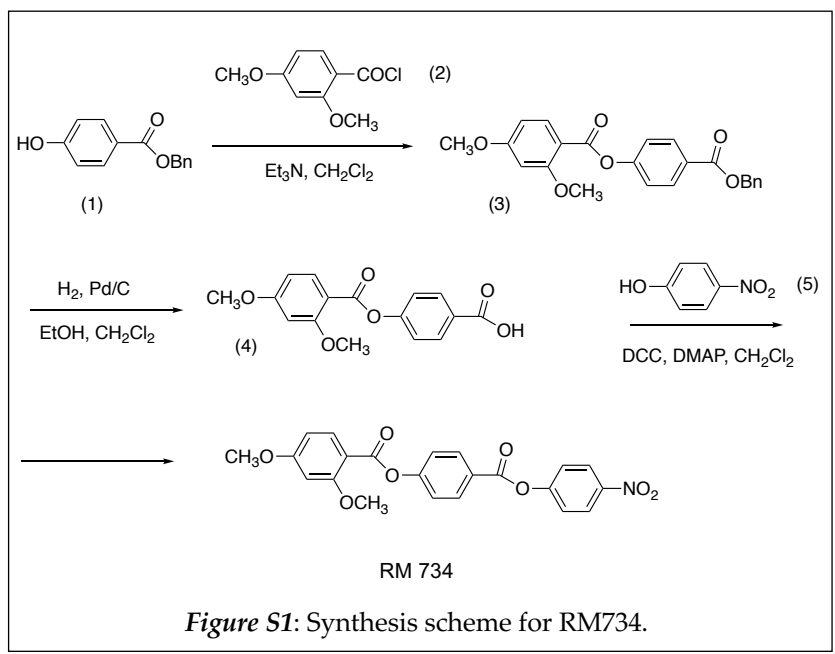

*Figure S1*: Synthesis scheme for RM734.

<u>4-[(benzyloxy)carbonyl]phenyl 2,4-dimethoxybenzoate (3)</u> – Benzyl 4-hydroxybenzoate (**1**) (5.69 g, 25 mmol) and 2,4-dimethoxybenzoyl chloride (**2**) (5.00 g, 25 mmol) were dissolved in $CH_2Cl_2$ (200 mL), and cooled to 0°C. Triethylamine (3.03 g, 30 mmol, 4.2 mL) was then added dropwise. The reaction mixture was stirred overnight at T = 25°C, then poured into a saturated aqueous solution of $NH_4Cl$ (200 mL) and extracted with dichloromethane. The combined organic layers were washed with water, brine, dried over $MgSO_4$, filtered and concentrated at reduced pressure. The crude product was purified by flash chromatography (silica gel, $CH_2Cl_2$/2% EtOH), to afford ester **2** as a white solid (9.4 g 96%, Mp. 89-92°C).

<u>4-(2,4-dimethoxybenzoyloxy)benzoic acid (4)</u> – A solution of the benzyl-protected carboxylic acid (**3**) (9.4g, 24 mmol) in $CH_2Cl_2$ (100 mL) and EtOH (100 mL) was first evacuated and purged with argon, then 10% Pd/C catalyst (2 g) was added. The argon atmosphere was replaced by hydrogen gas, and the reaction mixture was stirred at room temperature for two hours. Hydrogen was pumped out of the system and the flask was purged thoroughly with argon. The mixture was filtered through Celite and the solvents were removed under reduced pressure. This gave product as a white powder (6.99 g 96%).

<u>4-[(4-nitrophenoxy)carbonyl]phenyl 2,4-dimethoxybenzoate</u> – To a suspension of compound **4** (4.38 g, 15 mmol) and 4-nitro phenol (**5**) (2.02 g, 15 mmol) in $CH_2Cl_2$ (125 mL) was added DCC (6.00 g, 30 mmol) and trace of DMAP. The reaction mixture was stirred at room temperature for three days, then filtered, washed with water, 5% $CH_3COOH$, water, and brine, then dried over $MgSO_4$, filtered, and concentrated at reduced pressure. The resulting product was purified by flash chromatography (silica gel, $CH_2Cl_2$/2% EtOAc) followed by crystallization from $CH_3CN$. The product is a white crystalline solid (3.92 g 62%). 1H NMR (300 MHz, $CDCl_3$) ∂ ppm: 8.36 -8.25 (m, 4H), 8.10 (d, 1H) 7.46 -7.37 (m, 4H), 6.60-6.55 (m, 2H), 3.94 (s, 3H), 3.91 (s, 3H).  $^{13}C$ NMR (75 MHz, $CDCl_3$) ∂ ppm: 165.51, 163.73, 162.90, 162.64, 156.11, 155.82, 145.56, 134.74, 131.99, 125.69, 125.40, 122.76, 122.62, 110.53, 105.13, 99.16, 56.18, 55.75.

Differential scanning calorimetry measurements of RM734 are shown in *Fig. S2*.

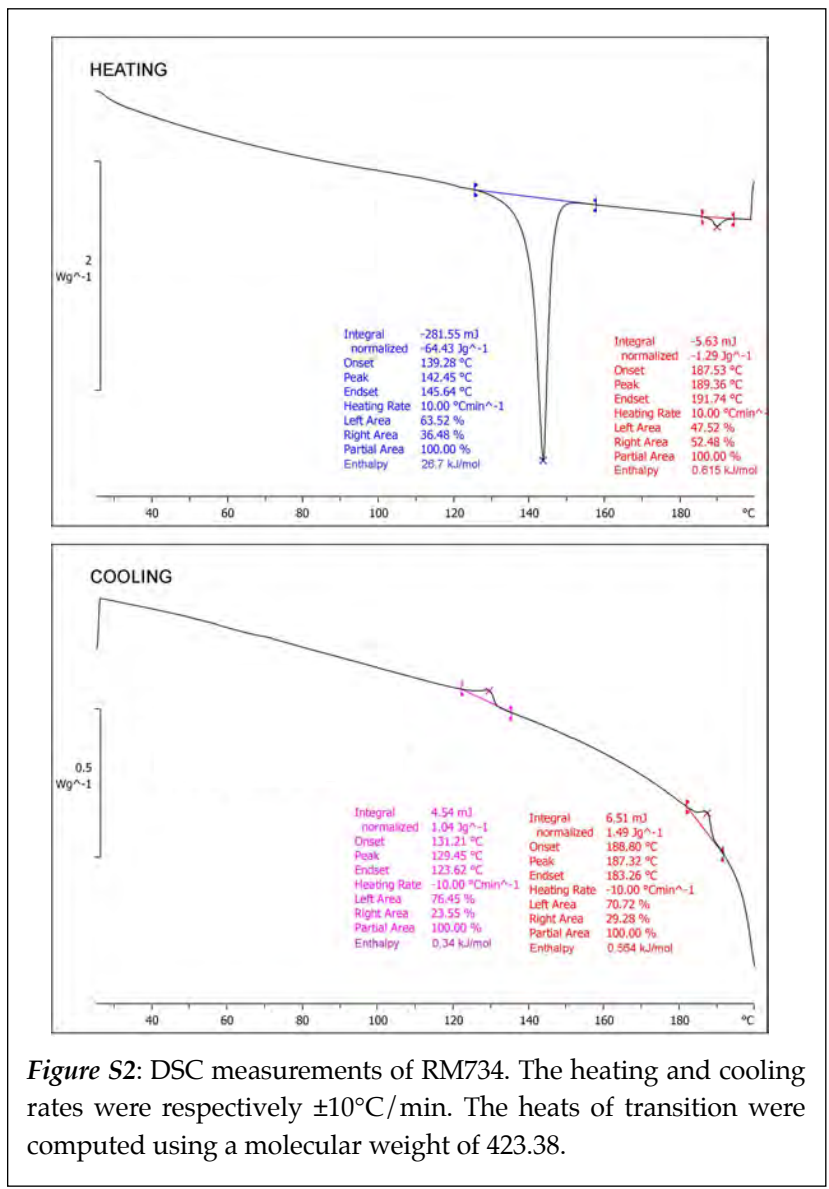

*Figure S2*: DSC measurements of RM734. The heating and cooling rates were respectively ±10°C/min. The heats of transition were computed using a molecular weight of 423.38.

*Methods* – <u>Electrooptic Measurement</u> – The liquid crystal phase sequence and phase transition temperatures were determined by polarized light microscopy using a Nikon Eclipse E400 POL microscope equipped with an Instec STC200 temperature-controlled hot stage. DSC was performed using a Mettler DSC823e differential scanning calorimeter. Electrooptic measurements were made using a Zeiss microscope, a Spectra Physics 105-1 HeNe laser, an EZ Digital FG-8002 function generator, and a Tektronix TDS 2014B oscilloscope.

<u>Polarization Measurement</u> – Polarization was measured by time integration of the current through two-terminal LC cells in which saturated polarization reversal was achieved. In-plane field cells, such as that in *Fig. S3A*, and the cylindrical, coaxial end-to-end electrode glass capillary cell in *Fig. S3B* were used.

Polarization and electrooptic measurements were carried out, in addition, in a 15 μm-thick Instec cell with two, 1 cm-long rectangular ITO electrodes patterned a distance 1 mm apart on one substrate, a geometry similar to that of the in-plane switching cell shown in *Fig. S3A*. The surfaces of the cell were treated with a planar alignment layer rubbed $87°$ from the electrode edges, the buffing direction $z$ thus making an angle of $3°$ with the in-plane electric field in the gap between the electrodes.

<u>Texture Cells</u> - Unless otherwise noted, the optical textures and domain structures of RM734 reported in the paper were observed in a $t = 11$ μm-thick LC Vision cell with a pair of 5 mm-long ITO strip electrodes spaced by $d = 1.04$ mm on one surface, giving an in-plane electric field. The plates had buffed polymer alignment normal to the electrode edges forming the gap, which gave weak anchoring alignment of the director in the direction of the applied field (*Fig. S4*).

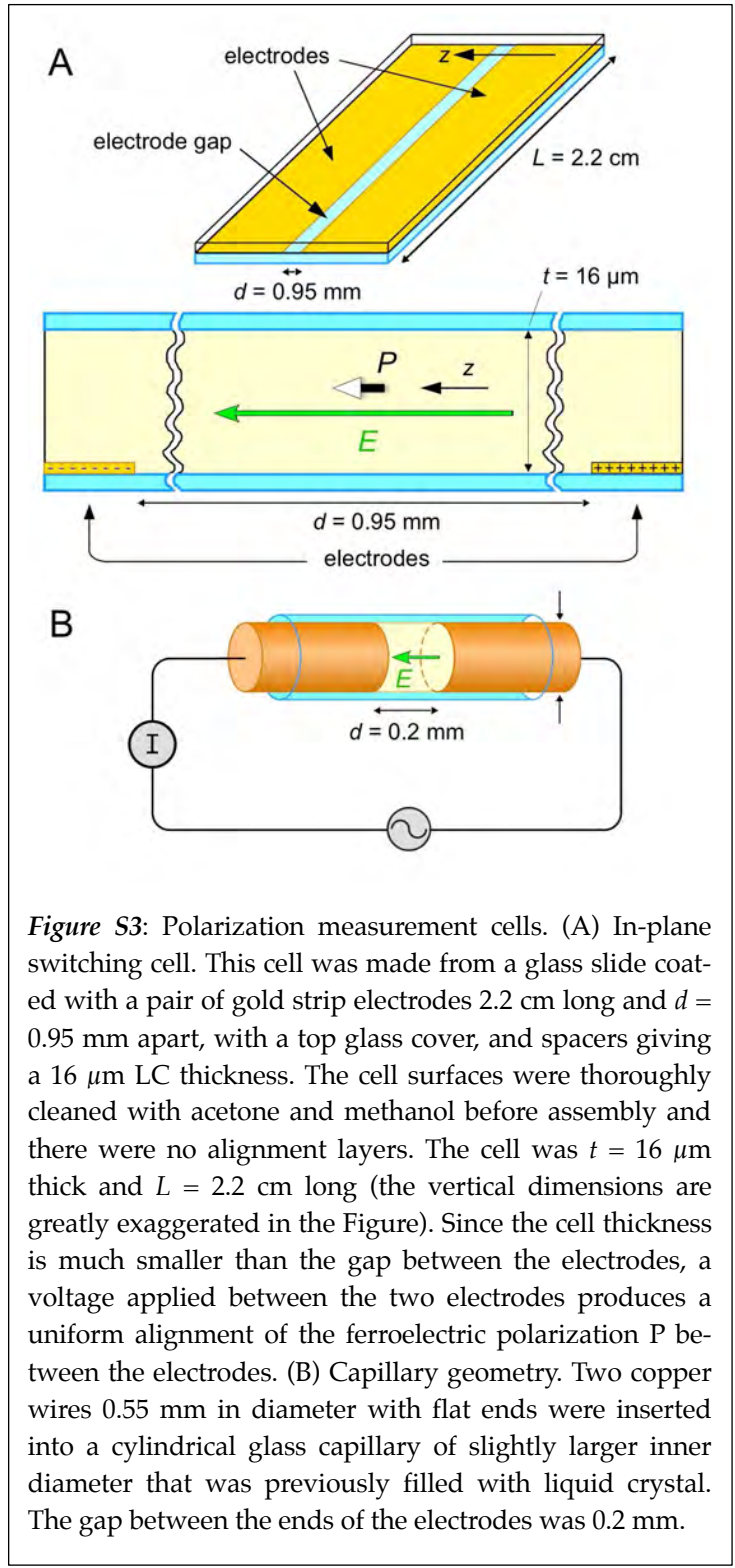

*Figure S3*: Polarization measurement cells. (A) In-plane switching cell. This cell was made from a glass slide coated with a pair of gold strip electrodes 2.2 cm long and $d = 0.95$ mm apart, with a top glass cover, and spacers giving a 16 μm LC thickness. The cell surfaces were thoroughly cleaned with acetone and methanol before assembly and there were no alignment layers. The cell was $t = 16$ μm thick and $L = 2.2$ cm long (the vertical dimensions are greatly exaggerated in the Figure). Since the cell thickness is much smaller than the gap between the electrodes, a voltage applied between the two electrodes produces a uniform alignment of the ferroelectric polarization P between the electrodes. (B) Capillary geometry. Two copper wires 0.55 mm in diameter with flat ends were inserted into a cylindrical glass capillary of slightly larger inner diameter that was previously filled with liquid crystal. The gap between the ends of the electrodes was 0.2 mm.

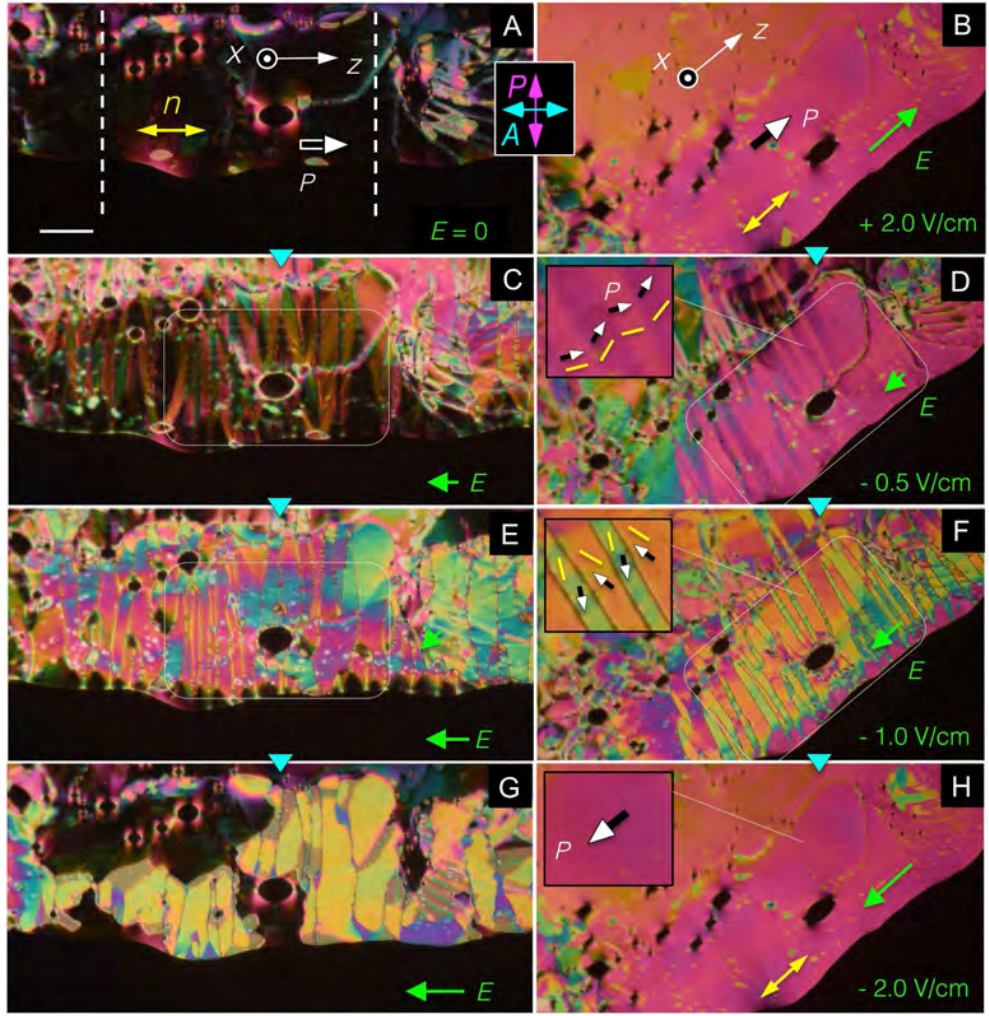

*Figure S4*: A larger area of the $t = 11$ μm test cell seen in *Figures 1* and *2* observed in DTLM, in orientations having the average director either along the polarizer/analyzer direction (**A, F–H**) or at 45° (**B–E**). The lower black area and the black ellipses are bubbles. (**A**) In the absence of applied field, the director *n*(*r*) is generally aligned along the rubbing direction *z*, producing good extinction. Experiments with small probe fields applied along the rubbing direction confirm that the average *P*(*r*) is directed toward the right. Vertical white lines show the electrode edges. (**B**) Rotating the cell reveals the pink birefringence color of the aligned domains (in the third-order Michel-Levy band). An applied field, $E = +2$ V/cm parallel to *P*, stabilizes the zero-field orientation. (**C,D**) Reversing the field to $E \sim -0.5$ V/cm begins the reorientation process, with bands in the texture corresponding to alternating rotation directions of the *n,P* couple. (**E,F**) At $E \sim -1$ V/cm the *n,P* couple has rotated by $\delta\varphi \sim \pm 90°$, forming regions of different signs of rotation separated by boundary walls. These regions are birefringent rather than extinguishing because *n*(*r*) twists non-uniformly across the thickness of the cell. (**G,H**) At $E \sim -2$ V/cm, the rotations of the *n,P* couple are nearly $+\pi$ or $-\pi$, with the regions of opposite rotation now separated by narrow $2\pi$ walls. This is immediately followed by a process in which domain boundaries nucleate holes whose growth leads to the elimination of the domain walls and the formation of large regions of uniform *n* and *P*, a state that is identical to that of (**A, B**) but with *P* reversed. This polarization reversal scenario takes place for fields $|E| < 2$ V/cm and is repeatable. Scale bar = 200 μm.

**SECTION S2 – HIGH POLARIZATION ELECTRO-OPTICAL, ELECTROSTATIC, AND ELASTIC EFFECTS**

*Director field response* – Given the very large polarization values of the $N_F$ phase, it is useful to summarize several of the relevant features of polar electro-optic, electrostatic, and elastic behavior, developed in the study of chiral smectic ferroelectric LCs, which can now be expected for the $N_F$: (***i***) <u>Polar twist Freedericksz transition</u> [3,4] - We take the uniform equilibrium state to have *P* along *z*, with *n(r)* and *P(r)* parallel to the (*y,z*) plane. In small applied fields, electrical torque on the director field $\tau_E = P \times E$ comes from the coupling of field to polarization. Applying this coupling in the description of the twist Freedericksz transition, the equation describing the azimuthal orientation field $\varphi(x)$ across the thickness of an otherwise uniform cell becomes $K_T\varphi_{xx} + PE\sin\varphi = 0$, with the threshold characteristics given above. (***ii***) <u>Boundary penetration</u> - Solving $K\varphi(z)_{zz} + PE\sin\varphi(z) = 0$ instead in the *y,z* plane, with the boundary condition $\varphi(z=0) = 0$, using the one-elastic constant (*K*) approximation, and applying an electric field to stabilize $\varphi(z) = 180°$ at large *z*, a $\pi$ reorientation wall is established in the LC near $z = 0$, given by $\varphi(z) = 4\tan^{-1}[1 - \exp(-z/\xi_E)]$. The ferroelectric field penetration length $\xi_E = \sqrt{K/PE}$ gives the approximate width of the wall [5], the distance that the effects of local orientational pinning such as the wall can penetrate into the neighboring LC, assuming the latter has *P* held in place by *E*. The penetration depth is $\xi_F \sim 1$ $\mu$m for an applied field $E = 1$ V/cm).

(***iii***) <u>Block polarization reorientation and expulsion of splay (splay-elastic stiffening)</u> [6,7,8,9,10,11,12,13] - Spatial variation of *P(r)* generally results in bulk and surface polarization charge density, given respectively by $\rho_P = \nabla \cdot P(r)$ and $\sigma_P = P_s \cdot s$. The electric field generated by the bulk charge opposes the bulk distortion of *P(r)* that caused it, producing a bulk energy $U_P = \frac{1}{2}\int dv \, \nabla \cdot P(r) \nabla \cdot P(r')[1 / |(r - r')|]$. Assuming a periodic transverse modulation $\delta P_y(r)$ of amplitude $P\delta n_y$ and wavevector $q_y$, so that $\nabla \cdot P(r) = \partial P_y(y)y = iq_yP_z\delta n_y$ in our geometry, we have an elastic energy density $U_{sp} = \frac{1}{2}[K_sq_y^2 + 4\pi P^2/\varepsilon]|\delta n_y|^2$, meaning that the polarization term will be dominant for $q_y < \pi\sqrt{2}/\xi_P$, where $\xi_P = \sqrt{\varepsilon K/P^2}$ is the polarization self-penetration length. Since for $P = 6$ $\mu$C/cm$^2$ we have $\xi_P \sim 0.1$ nm, this dominance will act down to molecular length scales. The result is that low-energy elastic distortions of the *n,P* couple allow only bend, with splay of *n(r)* and *P(r)* expelled from the bulk and confined to reorientation walls of characteristic width $\xi_P$. On the other hand, if we consider a longitudinal modulation $\delta P_z(y,z)$, the additional electrostatic free energy density will be $U_P = \frac{1}{2}[4\pi P^2 q_z^2/\varepsilon(q_z^2 + q_y^2)]|\delta P_z|^2$ [8].

(***iv***) <u>Field-step reorientational response</u> - The dynamics of polarization reorientation and the electro-optic (EO) response to changes in applied field are complex, depending on elastic, viscous, surface, and flow-induced torques in addition to that of the field. However, with the application of a large field step the electrical torques initially dominate and these determine the risetime of the optical response. The balance of field and viscous torques gives a characteristic reorientation risetime on the order of $\tau = \gamma_1/PE$, where $\gamma_1$ is the nematic rotational viscosity [4,14] (***Fig. 3C***). The risetime $\tau = \gamma_1/PE$ is $\sim 0.1\Delta t$, where $\Delta t$ is the reversal time (***Fig. 3C***), giving a value of $\gamma_1 \sim 0.1$ Pa sec, comparable to that of 5CB at $T = 25°$C.

(***v***) <u>The N–$N_F$ phase transition</u> is strongly affected by the polarization self-interaction, which suppresses the longitudinal modulation $\delta P_z$ of *P* [7], resulting in considerable anisotropy of the polarization fluctuations in the N phase, and rendering the transition mean-field, discussed in ***Sec. S3*** below.

*Charge screening* – All such polarization-based effects are reduced by free space charge, such as ionic impurities in the LC and its containing surfaces [9,10], ionization of the LC itself, and charge injected from the electrodes, all of which tend to screen the polarization charge. In SmC* FLC cells, when the polarization is small ($P < 20$ nC/cm$^2$) the bound polarization charge can be substantially screened but for large polarizations (P > 100 nC/cm$^2$) the free charge supply can be exhausted and polarization effects manifested. For the largest SmC* polarizations ($P$ ~ 800 nC/cm$^2$), the polarization charge is largely unscreened and the polarization effects are quite dramatic [15]. Thus, the SmC* FLC literature provides examples of screened and unscreened FLC behavior from which we can infer that the polarization charge of the N$_F$ phase is largely unscreened by ions. The DC conductivity of our RM734 sample is σ ~ 10$^{-7}$ (Ohm-cm)$^{-1}$, which is small enough to contribute little to the polarization measurement current (Fig. 3A) and not to affect the N-phase Freedericksz transition, but large compared to typical 5CB for example, which is σ ~ 10$^{-10}$ (Ohm-cm)$^{-1}$. Estimating the Debye length $\lambda_D$ using the ion density from σ gives $\lambda_D$ ~ 1mm, consistent with the observation that applied in-plane fields appear to be generally uniform and unscreened over millimeter dimension areas as in Fig. S4, so the Debye length must be this size or larger.

### *SECTION S3 – THE N–N$_F$ AND N$_F$–N PHASE TRANSITIONS*

*Ferroelectric domains* – As is evident in *Figs. 4,S5-S11*, the optical texture of the N phase in our planar-aligned cells is locally smooth. This texture roughens upon cooling toward the N$_F$ phase, developing a random pattern of micron-size domains extended along *n*, which, once in the N$_F$ phase, coarsen into a smooth texture very similar to that of the nematic but marked by a pattern of smooth lines, some of which extend for long distances generally parallel to *n* and others which form closed loops to make macroscopic, lens-shaped domains, as seen in *Figs. 1,S6,S7,S11*. The experiments presented in *Figs. 1,S5,S11* show that these lines demarcate areas of opposite polarization density, all formed in the absence of applied electric field. This remarkable evolution and the response of these domains to applied field are at the heart of the phenomenon of ferroelectricity, giving rise to its characteristic behaviors of macroscopic polar ordering and hysteretic dielectric behavior. In the weakly buffed cells studied here and in unbuffed cells, for thicknesses in the range 3 *μ*m < *t* < 15 *μ*m, the domain formation is similar to that reported here. Domain walls are generally extended along *n* and may have some characteristic spacing early in the coarsening process (*Fig. S9*), but toward the end of coarsening the spacing and loop size become very irregular - they appear to get pinned to the surfaces. The weak buffing in the *t* = 11 *μ*m cells used here did not exhibit an in-plane polar alignment preference.

Generally, the domain walls are visible because they have a width comparable to the cell thickness, and they appear to have an intra-wall director with a boundary condition that tends to keep *n* locally parallel to the wall (*Fig. 6A*). A resulting feature of these walls, evident in *Figs. 1,S5,S11*, is that wherever they run exactly parallel to *n(r)*, they lose optical contrast and can no longer be seen in DTLM. This means that at these places there is no reorientation of *n(r)* upon crossing the domain boundary: *n* is in the same direction inside and outside of the lens-shaped domains and through the wall, whereas *P* changes sign. We refer to these as pure polarization reversal (PPR) lines. PPR lines show that the differing states of the low-temperature phase found across domain walls differ only in the sign of *P*, and that *P* is therefore a suitable order parameter for describing the transition, *i.e.*, that the N$_F$ phase is a proper ferroelectric. PPR surfaces form upon cooling through the N–N$_F$ transition as the correlated volumes having predominately a particular sign of *P(r)* grow in size and the polarization fluctuations become slow and glassy. The resulting three-dimensional domains ultimately establish an equilibrium degree of order, encoded in the magnitude of *P*, and confine the transitions to the opposite sign of *P* in neighboring regions into PPR surfaces of a local structure that includes an internal nematic sheet, the central surface of the wall where *P* actually changes sign. In thin cells, these PPR walls have the appearance of lines but they are of course three-dimensional, with a structure controlled in part by the aligning surfaces of the cell. In a bulk 3D system, the PPR walls must form some sort of closed-shell structures.

With this observation, we now have two distinct kinds of defect lines in our cells where *P*(*r*) can change sign relative to the lab frame, **z**: the PPR lines in which *P*(*r*) flips by 180º but *n*(*r*) is unchanged, and the π–disclination (*Pn*D) lines of *Figs. 2* and *S6*, in which *P*(*r*) and *n*(*r*) rotate together, maintaining a fixed relative orientation.  Observations of the PPR lines show that they avoid making angles ψ relative to **z** larger that about 45º.  This is seen, for example, in the lens domains where the PPR line is forced to be a closed loop.  The line, rather than forming a continuously circular domain wall, where ψ would at two locations get to be 90º, instead jumps from ψ ~ -45º to ψ ~ +45º in order to complete the loop.

Similar behavior is also shown in *Fig. S6* for an extended line, where such a jump is maintained until ψ can be become sufficiently small without it.  π-PPR and π-*Pn*D lines can combine to make 2π walls in the uniquely ferroelectric way shown in *Fig. S7*.

*Modeling the $N$-$N_F$ phase transition* – The low-temperature phase of RM734 has been proposed to be a "splay nematic" phase [16,17,18], a splay-bend ground state of the sort initially described theoretically by Hinshaw et al. [19], in which the packing frustration of electrically and structurally polar (pear-shaped) molecules is relieved by the spontaneous adoption of local splay deformation, with 3D space filled by planar (x,z) sheets of splay ($\partial n_y / \partial y$) alternating in sign of splay and therefore also in sign of polarization (see Ref. [18], Fig. 11).  However, we find that: (*i*) the $N_F$ develops a large, macroscopic polarization as it grows in field-free through the transition from the N phase, and (*ii*) the ultimate polarized state at low temperature is effectively saturated, exhibiting no other significant birefringence or internal structural change upon cooling in the $N_F$ phase and with little response to large applied fields tending to increase *P* (i.e., *E* parallel to *P*).  These conditions appear to preclude any significant cancellation of *P* by opposed splay domains.

We propose that the ground state of the $N_F$ phase has, aside from fluctuations, uniform *P* and that it is a proper ferroelectric phase, where *P* is the principal order parameter and phase stability originates from a combination of short-range interactions that favor neighboring molecular long axes to be parallel and with the same polar orientation (*Fig. 1*), and long-range dipole-dipole interactions.  Under this condition, the N–$N_F$ transition would be viewed as Ising-like, with the N phase being a collection of molecules with long axes on average parallel to *n* but with disordered polar directions ($<p_i> = 0$), and the $N_F$ phase also having long axes on average parallel to *n* but with polar order $N<p_i> = P$.  Ignoring for the moment the dipole-dipole interactions, with only ferroelectric short-range interactions this transition would be predicted to be second-order with 3D critical exponents of the Ising universality class [20].

Mertelj et al. have observed paraelectric pretransitional effects, including the softening of the splay elastic constant, $K_s$, and a diverging dielectric anisotropy Δε in the N phase, that indicate growing polarization correlations, and flexoelectric coupling between these fluctuations and director splay as the transition is approached from above, observations which they have interpreted using a Landau-de Gennes model [17,18,21].  The free energy of Ref [18],   $f = ½ [\tau(T)(1 + \xi(T)^2 q^2)\delta P_z(q)^2 + K_s q_y^2 \delta n_y(q_y)^2] + \gamma q_y \delta n_y(q_y) \delta P_z(q_y)$, where $\xi(T)^2 = b/\tau(T)$, $\tau(T) \propto (T-T_c)/T_c$, and **q** = **q**$_z$ + **q**$_y$, includes Ornstein-Zernicke polarization fluctuations about $q = 0$ (no splay modulation), originating from short-range polar interactions,  and the flexoelectric coupling of $P_z$ to director splay.  This model successfully describes a mean field-like behavior of $K_s(\tau)$ and Δε(τ).  This free energy is consistent with the $N_F$ phase having uniform *P*, with a weak tendency for splay that is suppressed by the polarization charge effects discussed above, if the energy cost of the defects required to introduce splay modulation is included.

An additional pretransitional effect not yet adequately described is the striking anisotropy of the pretransitional and coarsening correlations in the fluctuations of *P* through the transition (*Fig. S8, S9*), with the domains becoming increasingly extended along *z*, until they end up as polar regions separated by PPR walls running largely parallel to *z*.  Given the significant anisotropies introduced into the $N_F$ phase by space charge effects, it is natural to consider that the polarization charge energy $U_P$ associated with electric dipole-dipole interactions may also affect the phase transition and pretransition fluctuations.  The resulting interaction energy combines short-ranged ferroelectric and long-ranged dipole-dipole forces.

The critical behavior of such systems has been studied extensively in an effort to understand certain magnetic materials that have short-range ferromagnetic exchange forces, but where the long-range dipolar interactions are also important [22,23,24].  In these systems, short-ranged interactions are included in a model Hamiltonian as nearest-neighbor Ising or Heisenberg-like, and the long-ranged interactions are calculated explicitly.  Renormalization group analysis shows that the long-range interactions make the magnetic correlations dipolar-anisotropic near the transition in the high temperature phase [25,26], extending them along *z* by strongly suppressing longitudinal charge density ($\partial P_z/\partial z$) fluctuations [22,23]. Specifically, starting with the free energy expression Eq. (1) from ref. [18] and adding the dipole-dipole interaction term $U_P$ from above, the structure factor for fluctuations in $P_z$ becomes $\langle P_z(q) P_z(q)^* \rangle = k_B T \chi(q)$ where $\chi(q) = [\tau(T)(1 + \xi(T)^2 q^2) + (2\pi/\varepsilon)(q_z/q)^2]^{-1}$. The dipole-dipole (third) term produces extended correlations that grow as $\xi(\tau)$ along *x* and *y* but as $\xi(\tau)^2$ along *z* [23], suppressing $\chi(q)$ for finite $q_z$ as is observed qualitatively from the image sequences of the textures upon passing through the phase transition, and from their optical Fourier transforms in ***Fig. S9***.  Because of this anisotropy, the correlation volume in this model grows in 3D as $V \sim (\tau)^4$ rather that the isotropic $V \sim \xi(\tau)^3$, reducing the upper marginal dimensionality of the transition to 3D, making the transition mean-field-like with logarithmic corrections, rather than fluctuation-dominated with 3D Ising universality [27].

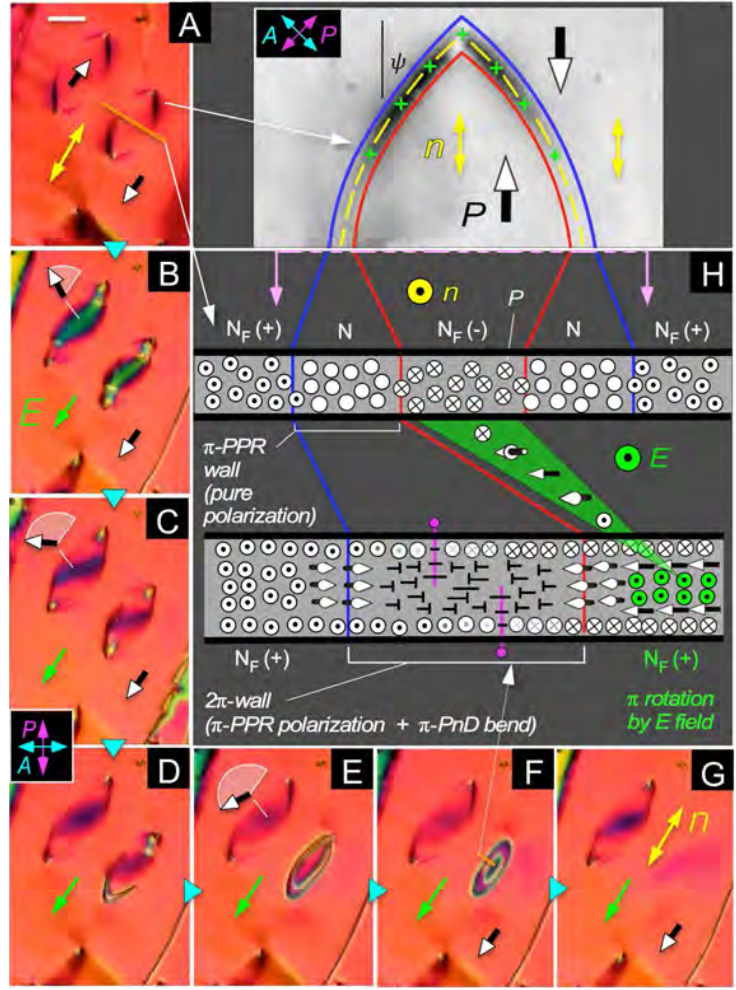

*Figure S5:* Step by step through the life of a lens-shaped $N_F$ domain in an applied reversal field increasing in the range $0 < E < 1$ V/mm. (**A**) Virgin domains before the field is applied. Yellow arrows show the director *n*, green the applied field *E*, and black the polarization, *P*. The director at the domain wall tends to follow the wall, otherwise *n* is uniform within each domain. The domain wall is invisible at the widest part of the lens, indicating that the wall does not involve reorientation of *n*, only reversal of the polarization – a pure polarization reversal (PPR) wall. The lenticular shape of the domain is due the interaction of *n* inside the wall with the surface buffing, making PPR lines stable only for $\psi <\sim 45°$. (**B**) In a small applied field ($E \sim 0.1$ V/mm), the domain interior reorients through an azimuthal angle $\varphi \sim 90°$. The area outside the domain, does not reorient, indicating that it is already aligned with the field. (**C**) Reorientation approaches 180°. (**D–F**) A reorientational disclination loop nucleates in the cell mid-plane to accommodate the induced 180° reorientation of *n*, and moves in response to the applied field to surround the second domain. (**F,G**) The domain shrinks as the disclination loop collapses. The first lenticular domain also responds to the field but maintains its shape, its boundary apparently trapped more strongly. (**H**) Cross-sectional sketches of the domain structure before and during field application. The initial domain boundary is a PPR polarization wall with no disclinations in *n*. In the region between $N_F(+)$ and $N_F(-)$, the system must pass through $P = 0$, *i.e.*, be nematic. The field-induced reorientation inside the disclination loop (green) adds an additional 180° rotation (a polarization-director disclination, *Pn*D) to the domain wall, as well as surface twist disclinations inside the loop which connect the bulk field-induced reorientation to the surfaces. The green swoop indicates the field-induced reorientation of the domain interior, causing the formation of a compound π-PPR / π-*Pn*D boundary to make a 2π wall. The reorientation on the surface is mediated by π twist disclinations (magenta). Scale bar = 100 μm.

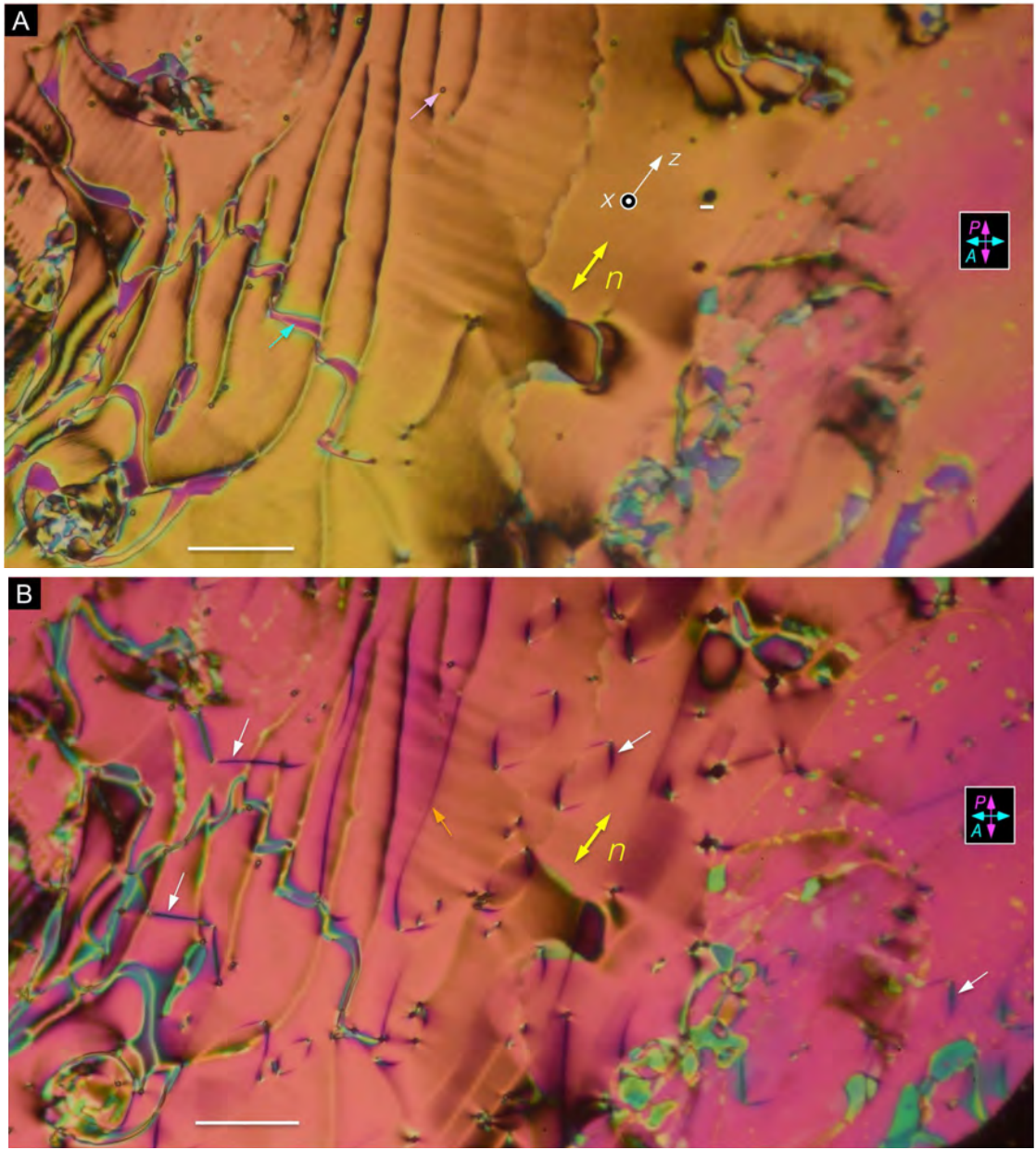

*Figure S6*: Typical DTLM images of the $t = 11$ μm, planar-aligned, in-plane field cell with $E = 0$. (**A**) N phase ($T = 140°$C). (**B**) $N_F$ phase ($T = 120°$C). The director ***n*** is generally along the buffing direction *z*, here making an angle of about 45° with the crossed polarizer and analyzer but giving good extinction when rotated so that ***n*** is along the polarizer or analyzer direction. The linear defects in the texture are π director disclinations and surface memory reorientations [28] trapped at the I–N transition. The director is generally uniform along *x*, the normal to the plates, except for a few twisted areas [cyan arrow in (**A**)]. Several 10 μm-diameter spacer particles are visible [pink arrow in (**A**)]. White arrows in (**B**) indicate typical Pure Polarization Reversal (PPR) lines, which upon cooling from the N phase, mediate the reversal of *P* in space; the orange arrow points to a PPR line running nearly along ***n***. The N and $N_F$ textures are locally smooth (optically featureless), except near the phase transition as illustrated in *Figures S8,9*. The distribution of ***P*** along +*z* and -*z* is roughly equal. Scale bar = 200 μm.

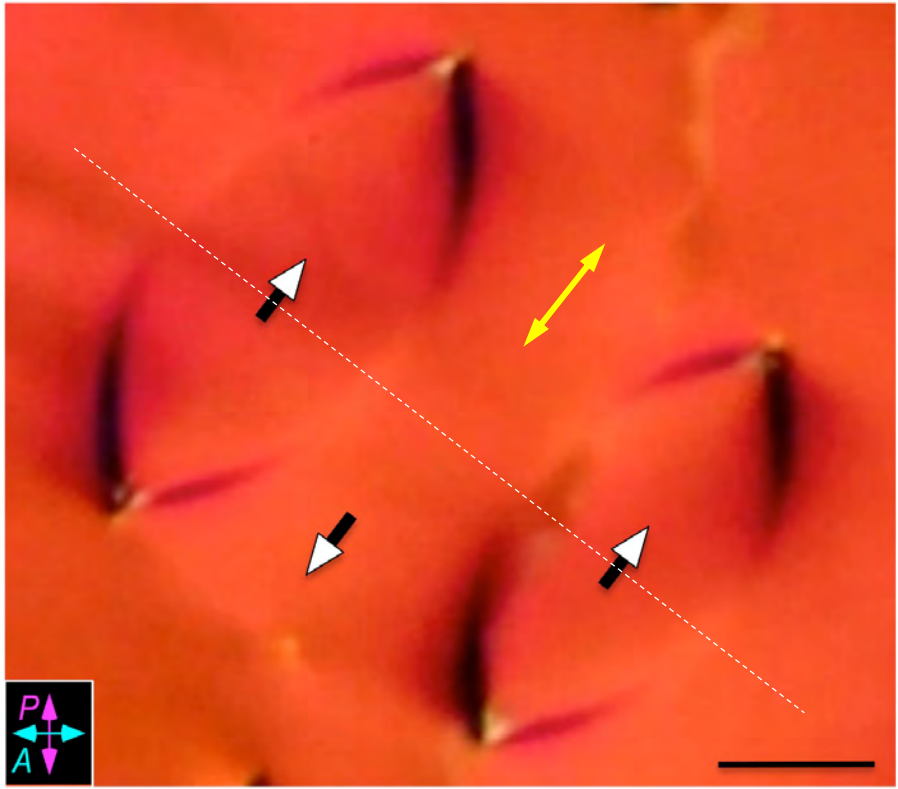

*Figure S7*: Magnified view of the polarization domains in *Fig. S6.* The domain boundaries become invisible when they are aligned parallel to the director field *n* (yellow). This means that the director field is uniform along the dashed white line and the domain walls are pure polarization reversal (PPR) in a sea of uniform ferroelectric nematic (*Fig. S5*). The domain walls become visible when they run at an angle with respect to *n*, because their internal director tends to be parallel to the wall. Thus the walls extinguish where they are vertical in the image, parallel to the incident optical polarization. $T = 120°C$. $t = 11$ μm. Scale bar = 50 μm.

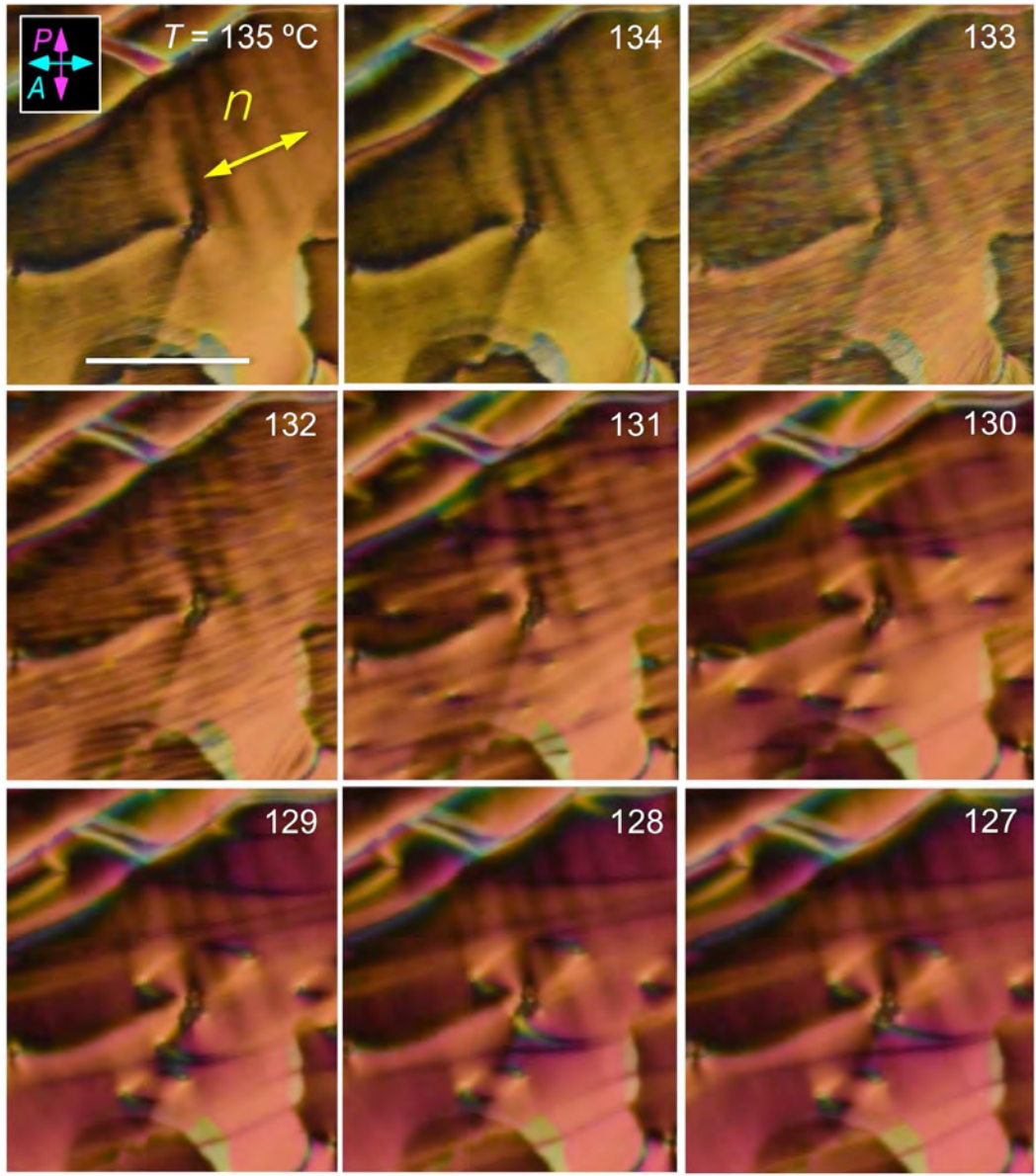

*Figure S8*: DTLM images obtained while cooling through the N–N$_F$ transition, showing anisotropic polar correlations growing in size in the N phase as the transition to the N$_F$ at T = 133°C is approached. This pattern of bright and dark domains exhibits dynamic fluctuations in the N phase. We propose that in the N phase this extreme anisotropy is a result of the suppression of fluctuations along $z$ of the magnitude of $P(r)$ by polarization charge (dipole-dipole) interactions. Below the transition, the texture is characterized by macroscopic ferroelectric domains bounded by pure polarization reversal (PPR) lines which move and coalesce to coarsen the texture. $t = 11$ μm. Scale bar = 200 μm.

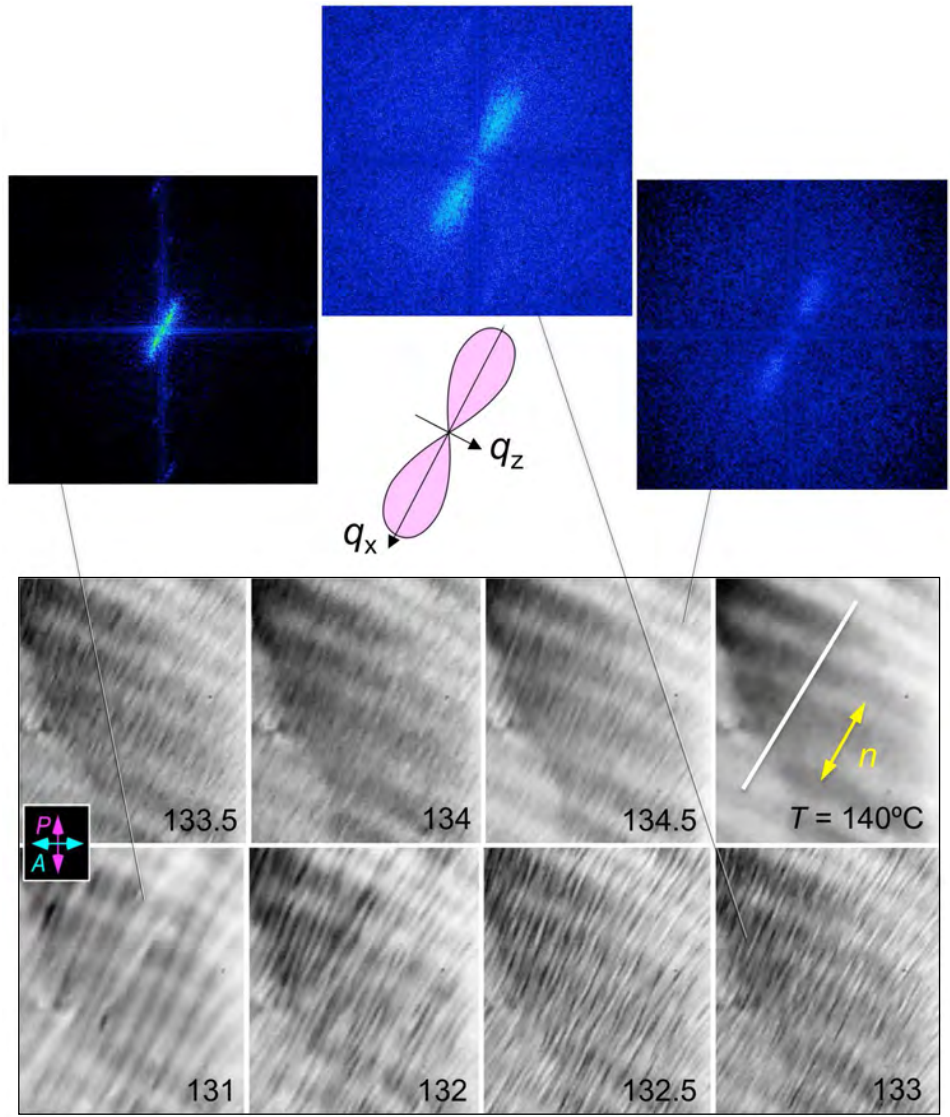

*Figure S9:* Grayscale DTLM images of a sample area of the RM734 texture and selected optical Fourier transforms near the N–N$_F$ phase transition. In the N phase (T < 133ºC), the images represent an optical resolution integration of the birefringent phase accumulated by the light upon traversing the sample thickness, $t$. The resulting averaging captures the extension of the domain shape along $n$, which gives the wing-shaped optical Fourier transform (OFT). The pink inset shows a contour of the intensity that would be scattered by polarization fluctuations in the x-ray region given by

$$\chi_P(q) \propto \langle P_z(q)P_z(q)^* \rangle \propto [\tau(T)(1 + \xi(T)^2 q^2) + (2\pi/\varepsilon)(q_z/q)^2]^{-1}$$

as discussed in the text. This susceptibility describes the anisotropy in the polarization fluctuations in the N phase generated by polarization charge interactions. The favorable qualitative comparison with the $T = 133$ºC OFT suggests that charge stabilization is an important factor governing fluctuations in the N phase. In the N$_F$ phase, PPR lines (domain walls) extend long distances along $n$, producing sharp intensity bands in the Fourier transform image that are normal to $n$. These textures suggest that resonant and non-resonant SAXS and WAXS could be effective probes of $\chi_P(q)$, as well as of the transverse fluctuations of $P(r)$. $t = 11$ μm. Scale bar = 200 μm.

***N–N$_F$ phase coexistence upon melting the N$_F$*** – Slow heating of RM734 through the N$_F$–N transition results in a distinctive melting process in which the nematic appears as extended, flexible filaments in a background of N$_F$. Examples of these filaments are shown in ***Fig. S10***, where a nematic front is moving in from the right as the temperature increases slowly. The filaments, which have diameters in the range 2 µm < $D_f$ < 10 µm, grow in from the right, where their density is largest. The nematic filaments penetrate the N$_F$ phase but do not otherwise disturb it, with the N$_F$ maintaining its alignment and birefringence independent of any nearby filaments. The nematic alignment inside each filament appears to be uniform, with the director oriented parallel to the filament axis, so that the filaments can generally be seen with high contrast, except where they are aligned parallel to the N$_F$ background field. As with the domain walls in the N$_F$ phase, the nematic filaments lose optical contrast when oriented parallel to the director field of the background N$_F$ phase. The orientation of the filaments does not appear to be strongly coupled to the N$_F$ director field and varies greatly across the sample. These filaments show that the N and N$_F$ phases can coexist as distinct states, at least meta-stably.

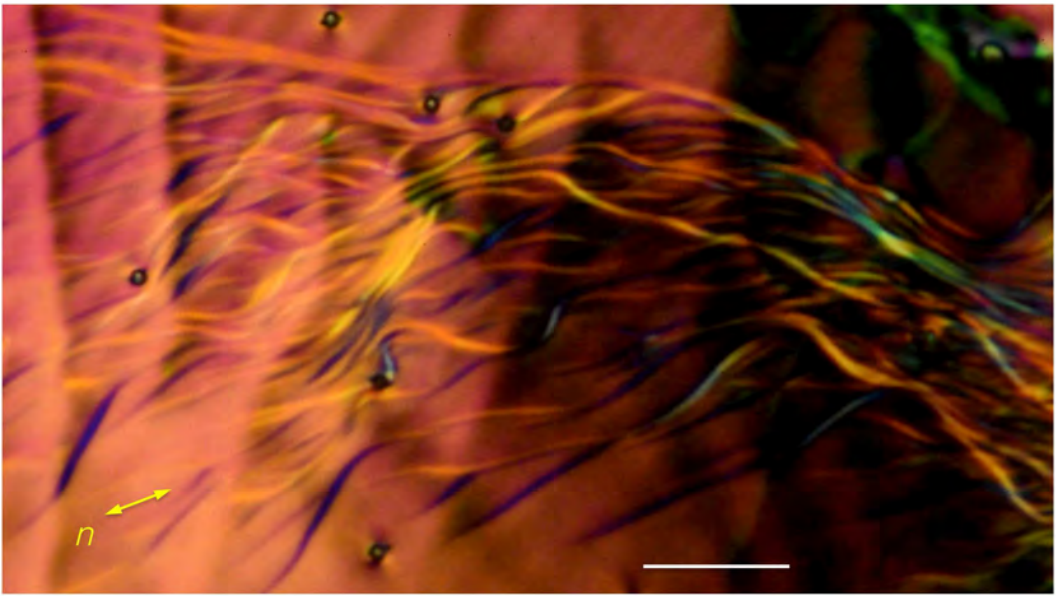

***Figure S10***: Filamentous domains of the N phase moving into the N$_F$ phase upon heating through the N–N$_F$ transition near $T = 130$ °C in a $t = 11$ µm cell. The N filaments grow in from the right. The filaments lose their optical contrast when they are oriented parallel to the director of the background (orange) N$_F$ phase, indicating that, as in the PPR walls, the director inside the filaments is oriented parallel to the filament boundary. This phase coexistence is evidence that the N–N$_F$ transition is first order. Scale bar = 100 µm.

*SECTION S4 – COMBINING PPR AND PnD DOMAIN WALLS*

The combination of π-PPR and π-*Pn*D lines creates a novel hybrid defect line unique to the $N_F$ phase. *Fig. S11A* shows two PPR lines separating domains of opposite *P*, with a small field beginning to reverse the central domain. The PPR lines are barely visible, as noted in *Fig. S7*, but increased field strength reverses the central domain, making it visible and leaving unswitched (green) sheets $\sim\xi_P$ thick at the surfaces. Then, at sufficiently high field, in a process analogous to the "order reconstruction" of confined splay in a nematic [29,30], pink holes open up in these sheets (*Fig. S11B*). This nucleation and the resulting movement of π-*Pn*D lines to the boundaries of the central domain, leave it completely pink, i.e., uniformly switched with *P* now along *E*, and now identical in structure to the outer area, separated from it by the hybrid π-PPR/ π-*Pn*D 2π-walls, now visible because of their *Pn*D component. The cross-sectional structure in *Fig. S11C* shows the 360° reorientation of *P* on passing through the wall and that in an applied field, such walls have energy in excess of that of the surrounding areas. This energy difference increases with field, which confines the 180° *Pn*D reorientation of the wall to a central sheet of width $\xi_E = \sqrt{K/PE}$, a high energy director configuration in a sea of uniform orientation. At sufficiently high field, "order reconstruction" holes [29,30] nucleate in the sheet, inducing gaps in the wall such that it recedes (*Fig. S11D-G*). This process is one-way: once the wall recedes, the area vacated shows no memory of it when the field is reversed. As a result, the PPR lines in a newly cooled cell tend to disappear by nucleation and movement of gaps when exposed to a changing field. The comparisons above of polar vs. dielectric response to field also apply to this process.

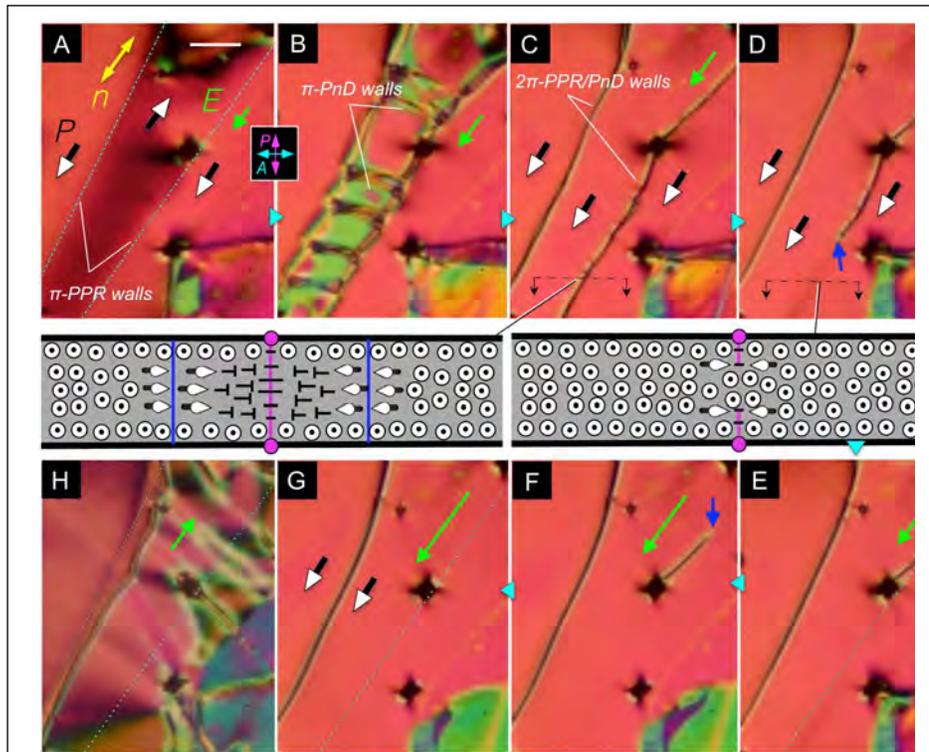

*Figure S11*: Polarization-reversal and hysteresis: (**A**) Domains with opposite polarization grown in at the N–$N_F$ transition (central domain up, outer domains down) separated by pure polarization reversal (PPR) π-walls. (**B,C**) Field-induced reversal of the central domain by bulk and surface reorientations, leaving it separated from the outer domains by composite 2π-PPR/*Pn*D walls. (**D–G**) Gap nucleation in, and shrinking of, 2π-PPR/*Pn*D walls, induced by increased field. Since the walls have higher energy than the uniform state that replaces them, they are squeezed out by the field (blue arrows) and do not regrow upon decreasing the field, a hysteretic response. (**H**) A second field reversal produces a bend instability in the director field like that shown in *Figure 5*. The sections in **C** and **D** show the 2π-PPR/*Pn*D wall structure and its squeezed-out remnant respectively. Scale bar = 100 *μ*m.

## SECTION S5 – SPLAY IN THE N AND $N_F$ TEXTURES

The evolution of the cell upon cooling into the $N_F$ phase, ending, apart from the presence of the domain walls in *Fig. 1*, in textures that are optically smooth with close to the same birefringence as the nematic, suggests that there are no sub-resolution orientational modulations or non-uniformities that would reduce the birefringence of the equilibrium $n(r),P(r)$ fields of the $N_F$ phase below that of the N phase. However, the polar symmetry of the $N_F$ phase implies a polar orientational distribution of the molecules, which, because of the lack of reflection symmetry of their steric shape, should generate a tendency for director splay in the $N_F$ phase that is correlated with the direction of $P(r)$, like that of flower stems in a bouquet, with $P(r)$ indicating the growth direction [31]. This effect has been observed in calamitic chiral smectics [32], and in polar-ordered smectic bent-core B2 phases that are ferroelectric ($SmC_SP_F$) [33]. The latter break up into structures of linear splay domains, an instability which generates the B7 phases, for example [34]. Such structures appear only above a threshold value of preferred splay amplitude because filling 3D space with splay always requires some sort of non-splayed or oppositely-splayed defects that cost extra energy [19,35]. Below this threshold, the splay is simply frustrated in domains of uniform orientation. In some situations, however, splay is required, such as in the $n(r),P(r)$ field around the air bubbles in *Fig. 5C* and *Fig. S12*, and around and inside the *P*-reversal domains in *Fig. 1D* and *Fig. S5*. Here different signs of polarization splay are required to make the two 180° wedge defects on opposite sides of the bubble (red dots), and at opposite ends of the polarization domains. By symmetry, these opposing structures must be different. However, they are in fact very similar in appearance, implying that the preference for a particular sign of splay does not significantly affect their structure. The $N_F$ phase therefore appears to have large polarization but a weak tendency for splay, indicating a small splay flexoelectric coefficient [36] that is perhaps a result of its molecular-scale organization, discussed in the text and in *Section S10*.

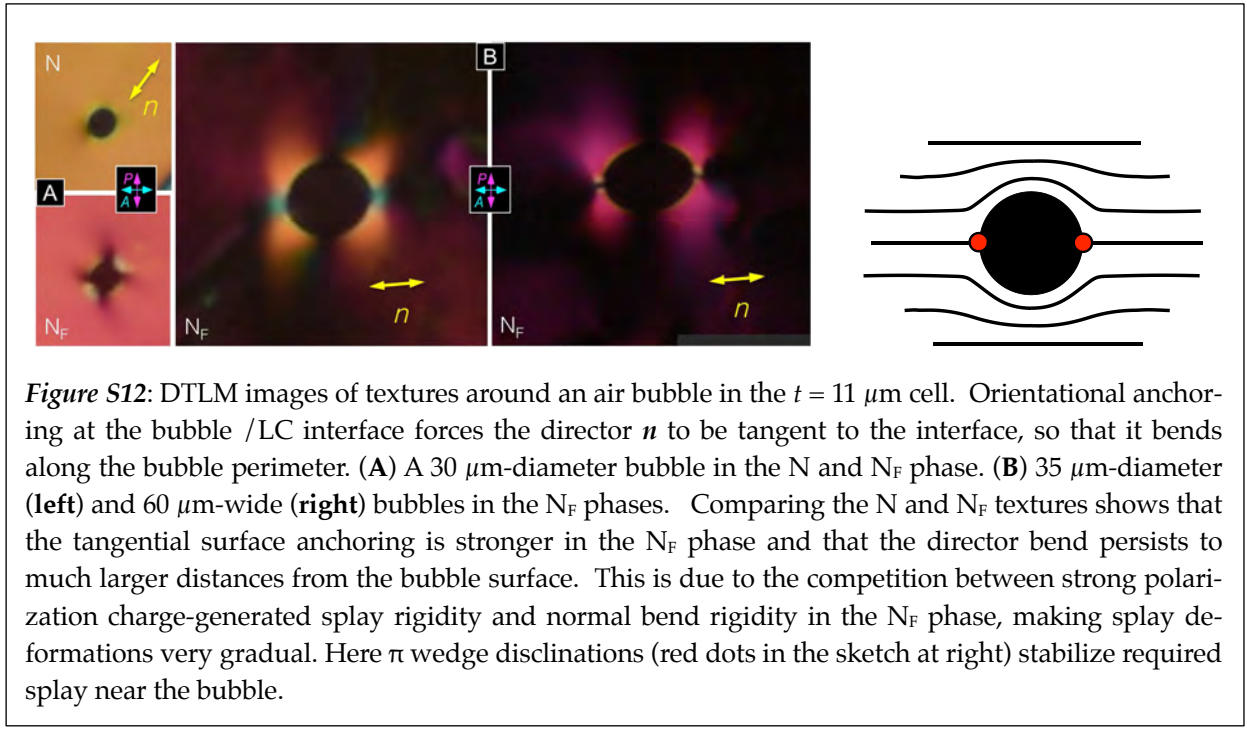

*Figure S12*: DTLM images of textures around an air bubble in the $t = 11$ µm cell. Orientational anchoring at the bubble /LC interface forces the director *n* to be tangent to the interface, so that it bends along the bubble perimeter. (**A**) A 30 µm-diameter bubble in the N and $N_F$ phase. (**B**) 35 µm-diameter (**left**) and 60 µm-wide (**right**) bubbles in the $N_F$ phases. Comparing the N and $N_F$ textures shows that the tangential surface anchoring is stronger in the $N_F$ phase and that the director bend persists to much larger distances from the bubble surface. This is due to the competition between strong polarization charge-generated splay rigidity and normal bend rigidity in the $N_F$ phase, making splay deformations very gradual. Here π wedge disclinations (red dots in the sketch at right) stabilize required splay near the bubble.

***SECTION S6**– S<small>ANDWICH</small> C<small>ELL</small> E<small>LECTRO</small>-O<small>PTICS</small> & <small>THE</small> S<small>PLAY</small>-B<small>END</small> F<small>REEDERICKSZ</small> T<small>RANSITION</small>*

The $N_F$ phase exhibits characteristic high-polarization ferroelectric LC behavior, illustrated in ***Fig. S13***, which shows a sequence of DTLM images of RM734 in a conventional, transparent capacitor cell $t = 4.5$ $\mu$m thick, with antiparallel-buffed polyimide-coated ITO electrodes that give unidirectional planar alignment with a few degrees of pretilt. The sample temperature is held near the N–$N_F$ transition, a small, in-plane temperature gradient ensuring that part of the sample is in the $N_F$ phase while the rest is in the N phase, with a wide transition region (demarcated by white and black dashed lines in ***Fig. S13A***) where the two phases form overlapping wedges (sketched in ***Fig. 8C***) and the birefringence is lower. In the N phase, the LC responds dielectrically to an applied 1 kHz triangle wave voltage of amplitude $V_p$, resulting in a decrease of birefringence $\Delta n$, beginning at $V_p \sim 1.5$ V. For $V_p \sim 3$ V, the birefringent optical retardance $\Delta nt$ is reduced from 900 nm to 300 nm [37], corresponding to $\theta(V_p)$ increasing from $\theta \sim 0°$ to $\theta \sim 54°$, a substantial rotation of ***n*** toward the field direction. However, the net birefringence increases continuously through the wedge region as the thickness of the N layer decreases (***Fig. S13C***). In contrast, the part of the cell in the $N_F$ phase exhibits no observable change in birefringence in this voltage range. This behavior can be understood on the basis of the "block polarization" model developed to explain V-shaped switching in high-polarization FLCs [15,38,39]. In this model, polarization charge induced by reorientation of ***P*** completely cancels $E$ in the LC and the polarization direction is electrostatically controlled. Insulating layers without reorienting polarization at the LC/glass interfaces such as the polymer alignment layer, charge depletion in the ITO, and bound polarization at the surface, are accounted for as capacitive elements with a net capacitance per unit area, $C$. With these assumptions, the orientation of the polarization field as a function of applied voltage $V$ is given by $\sin\theta(V) = V/V_{sat}$, where $V_{sat} = P/C$ [15,38,39], showing that when $P$ is large, a large applied voltage is required to achieve substantial reorientation of ***n,P***. This $N_F$ cell can be compared with the bent-core $SmAP_F$ ITO electro-optic cell of Ref [15], where a saturation voltage $V_{sat} = 15$ V was measured for material with $P = 0.85$ $\mu$C/cm$^2$, leading us to expect $V_{sat} \sim 100$ V for LC in the $N_F$ phase with $P \sim 6$ $\mu$C/cm$^2$ in a cell with 50 nm-thick insulating layers with $\varepsilon = 5$ at the two cell boundaries. It is therefore not surprising that the $N_F$ phase shows very little response in the applied voltage range explored here. In a further experiment, we cooled through the N–$N_F$ transition while continuously applying a $V_p \sim 16$ V triangle wave, observing Freedericksz-type reorientation to a homeotropic state in the N phase and the reimposition of planar alignment by polarization stabilization in the $N_F$ phase (see ***Fig. S14***).

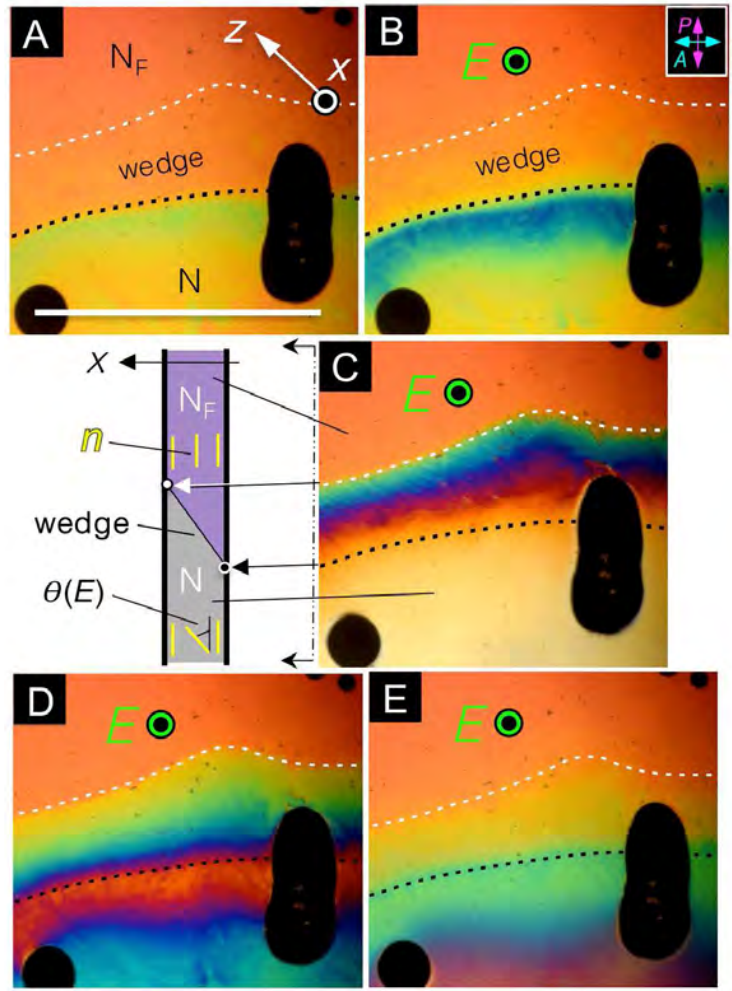

*Figure S13*: Splay-bend Freedericksz transition of RM734. The LC is in a $t = 4.5$ μm-thick transparent capacitor cell with polyimide-coated ITO electrodes antiparallel buffed along $z$ giving unidirectional planar alignment with a few degrees of pretilt. The temperature increases by several degrees from top to bottom of the image, creating a wedge-like interface between the $N_F$ and N phases. (**A**) Planar aligned RM734 with no field applied. The director away from the phase boundary is oriented essentially parallel to the glass ($\theta \approx 0°$). (**B**) In a 1 kHz triangle wave voltage of amplitude $V_P \sim 1$ V, the director begins to align with the field, lowering $\Delta n$. (**C**) With $V_P \sim 3$ V, $\theta(V_P)$ is increased to $\theta \approx 54°$ in the lower N part of the cell. The $N_F$ region, in contrast, shows no change in $\Delta n$ with field. The change in birefringence across the wedge reflects the different director configurations obtaining in the N and $N_F$ phase tongues. (**D**) The return to planar alignment ($\theta = 0°$) upon reducing the field shows some hysteresis, with the region of the N phase nearest the $N_F$ retaining a slightly larger tilt $\theta$. (**E**) Texture on reducing the voltage to $V_P \sim 1$ V, as in (**B**). Scale bar = 1 mm. The black regions are air bubbles.

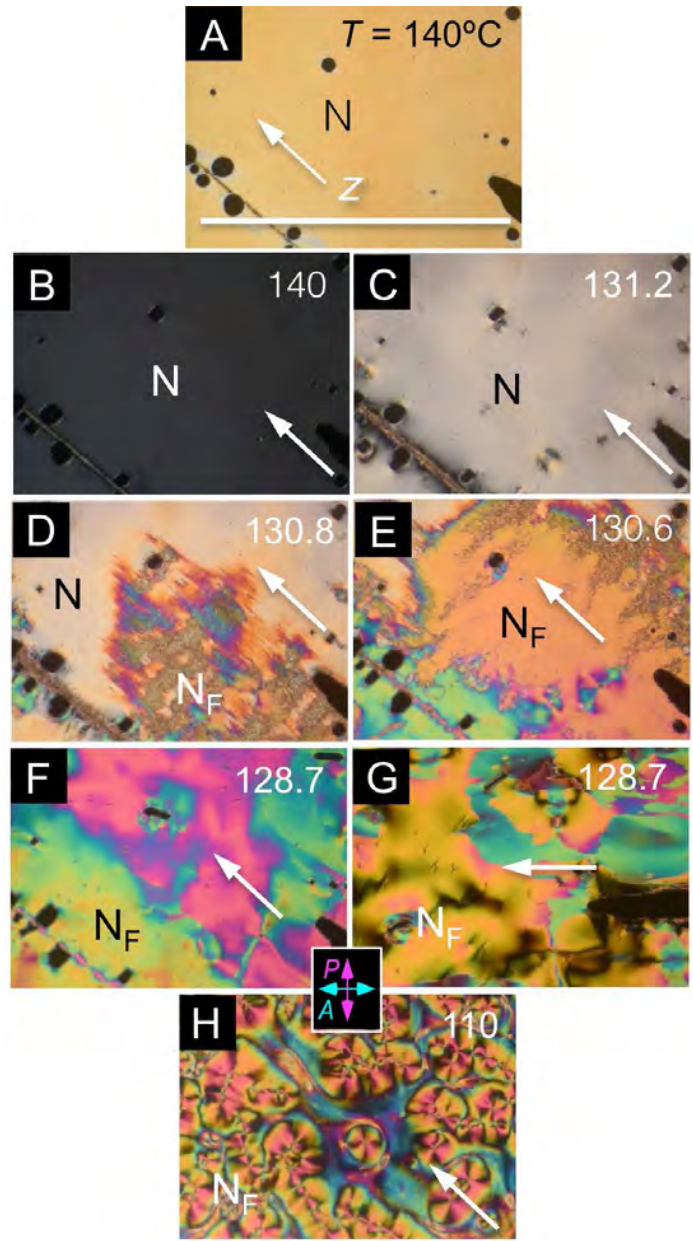

*Figure S14*: Splay-bend Freedericksz texture cooling sequence of RM734 in the ITO sandwich cell of *Fig. 8*. (**A**) Planar-aligned, with no field applied. (**B–H**) This cell is cooled through the N–N$_F$ transition while continuously applying a $V_P$ ~ 16 V, 500 Hz triangle wave, well above the Freedericksz threshold in the N phase. (**B**) The N phase with the field applied starts at $\theta$ ~ 85° (above threshold and nearly homeotropic) at $T = 140$°C but (**C**) on approaching the transition, $\theta$ decreases to $\theta$ ~ 80° near $T = 131$°C, where (**D**) the N$_F$ phase comes in via irregular domain boundaries that (**E**) anneal into an ordered $\theta$ ~ 0° planar-aligned N$_F$ domain (orange birefringence color). However, (**F–G**) as $T$ is lowered to $T = 120$°C this field induces a defected pattern of increased $\theta$, a birefringent texture that is permanently written when the field is removed. This process must involve breakdown of the interfacial capacitors discussed in the text. Texture (**H**) is obtained with a $V_P$ ~ 16 V, 1 Hz triangle wave. Scale bar = 1 mm.

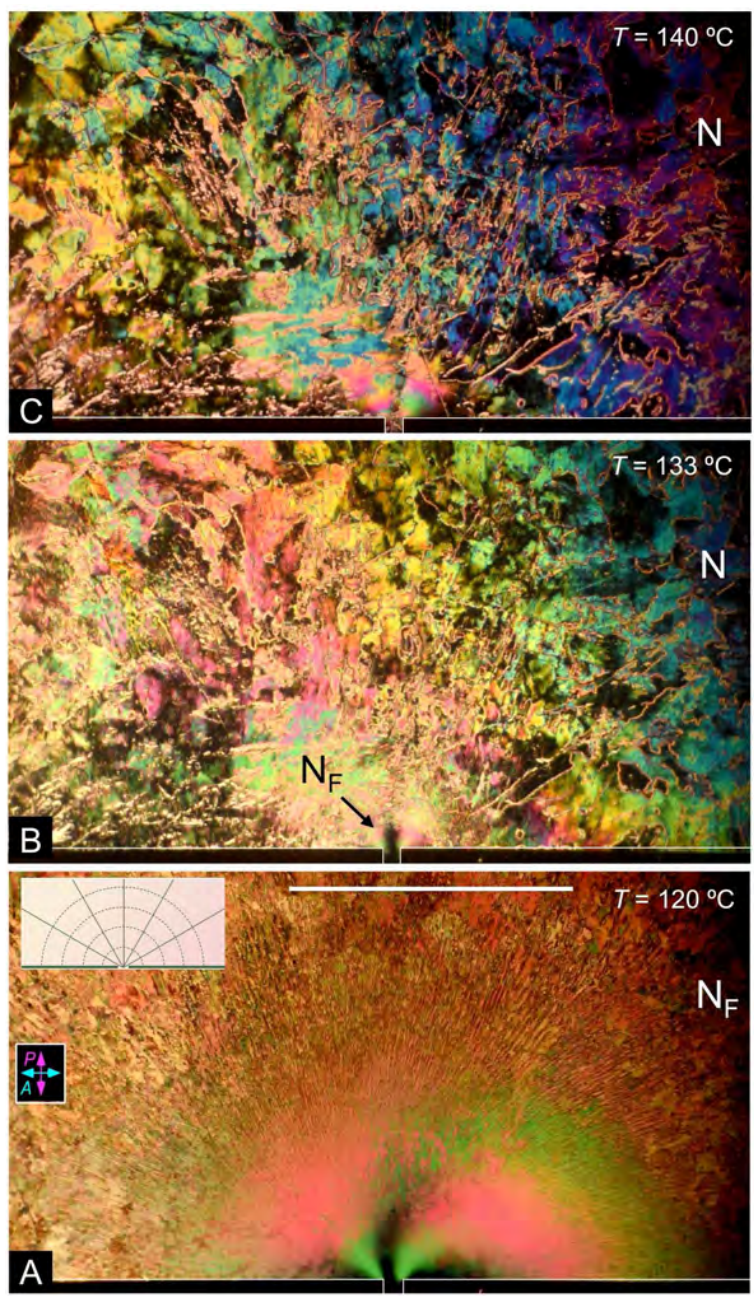

*Figure S15*: Field-induced flow in RM734 vs. temperature. DTLM images of a $t = 10$ μm-thick planar - aligned cell of RM734 between untreated glass plates. The black bars at the bottom of each image are two evaporated gold electrodes on one of the plates, separated by a $d = 60$ μm gap, outlined in white for clarity. A $V_P = 3$ V peak 0.1 Hz square-wave voltage is applied to the electrodes in all images. Only the upper portion of the electrodes and cell are shown. The cell is shown at three different temperatures: (**A**) $T = 120°C$ in the $N_F$ phase. The applied voltage drives a pattern of defect motion and fluid flow over the entire image, with the defect velocity $v(r)$ vectorially parallel to the applied field, $E(r)$, on circular arcs centered on the electrode gap. The graph shows the *E*-field lines (dashed) obtained by solving the Laplace equation. This image is taken at the instant of field reversal, where the resulting polarization reversal generates a periodic array of bend domain walls normal to the director, as in *Fig. 5A*, radial in this case (inset solid lines). (**B**) $T = 133$ °C. Most of the cell area is in the N phase and not flowing. The region near the electrode gap where the field is the highest is the last to remain in the $N_F$ and is still flowing. (**C**) T = 140°C. The LC is entirely in the N phase, with a Schlieren texture, and there is no observable flow, showing that the defect motion and flow are features of the $N_F$ phase (*Fig. 6*). Scale bar = 1 mm. See also *Fig. S16*.

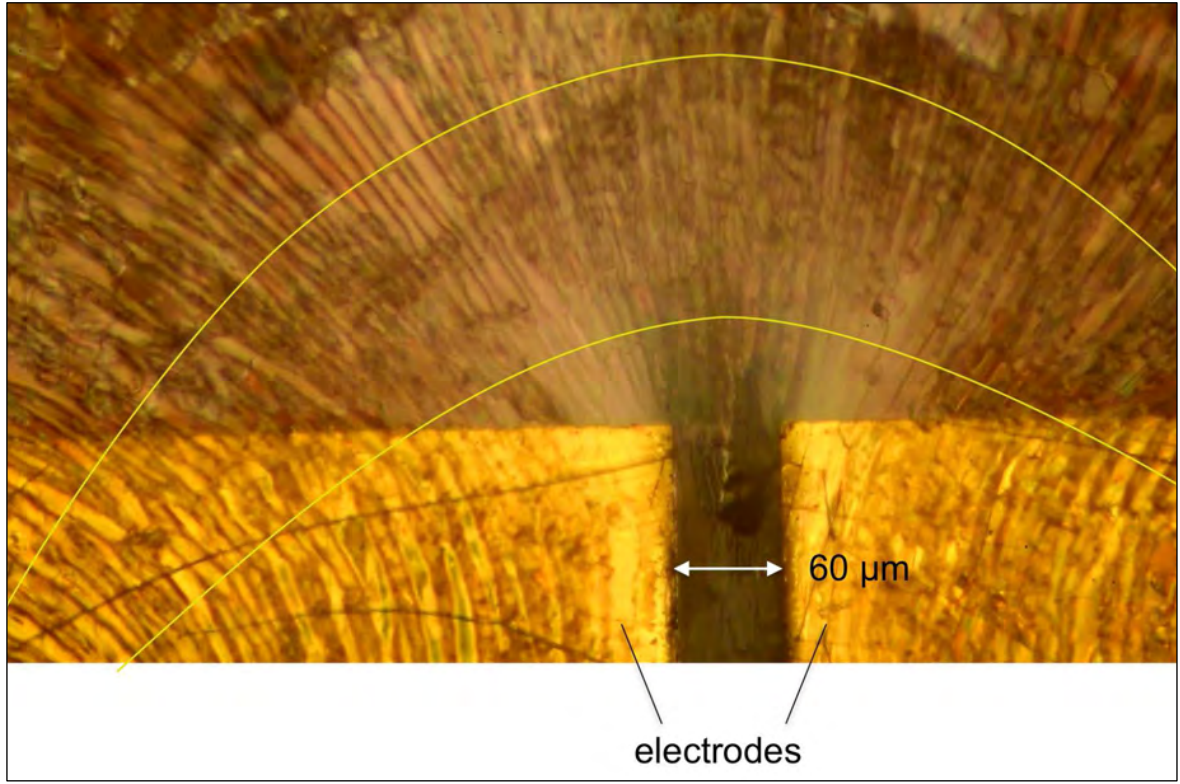

***Figure S16***: Reflected light microscope image of the cell in ***Fig. S15***, a $t = 10$ μm-thick planar-aligned cell of RM734 between untreated glass plates, with two evaporated gold electrodes separated by a $d = 60$ μm gap on one of the plates. This image was captured at $T = 120$ °C at the instant of applied field reversal, where the resulting polarization reversal generates a periodic array of bend domain walls locally normal to the reversing ***E*** and to the director field ***n(r)***, as in ***Fig. 5A***. At the top of the image, the lines are approximately radial as in ***Fig. S15A***. This image shows a remarkable feature of the $N_F$ electro-optics: even the weak applied field that penetrates over the electrodes is strong enough to produce polarization reversal in the LC. The yellow lines are along the director field.

## SECTION S8 – VARIATION OF UNIAXIAL BIREFRINGENCE WITH TEMPERATURE

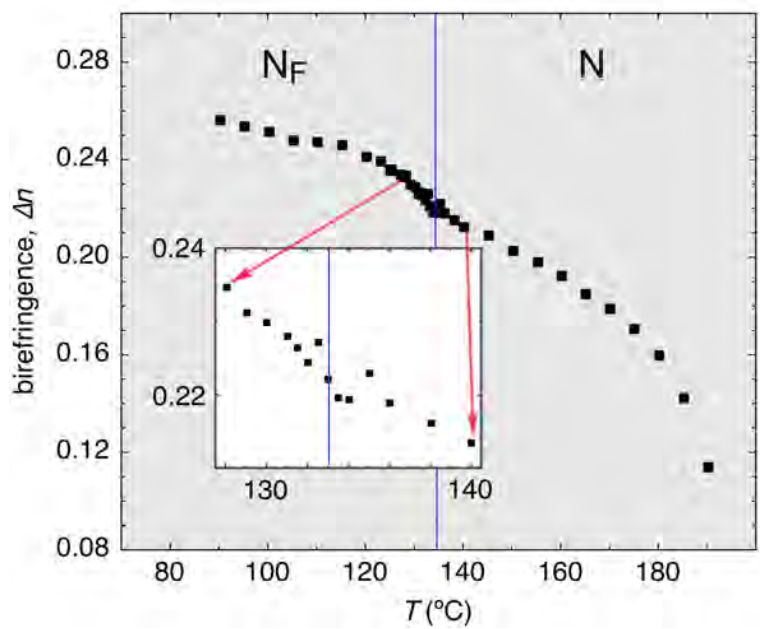

*Figure S17*: Birefringence of RM734, measured in a $t = 3.5$ μm-thick, planar-aligned cell using a Berek compensator.

## SECTION S9 – ATOMISTIC SIMULATION METHODOLOGY

*Force Field* – All molecular dynamics (MD) simulations were conducted using the Atomistic Polarizable Potentials for Liquids, Electrolytes and Polymers (APPLE&P) force field [40,41]. The parameters for atomic polarizabilities and repulsion-dispersion interactions were taken from the APPLE&P database without modification, while atomic charges were fitted to reproduce the electrostatic field around all of the molecular fragments as obtained from MP2/aug-cc-pVDZ quantum chemistry calculations using Gaussian 16 software [42,43]. The parameters for missing dihedral potentials were obtained by fitting conformational energy scans obtained from DFT calculations at the M052X/aug-cc-pVDZ level of theory [44]. A non-polarizable version of the force field was also used, with the atomic polarizabilities set to zero and all other parameters kept the same as in the polarizable version.

*Simulation Parameters* – The simulations were carried out using the WMI-MD simulation package (http://www.wasatchmolecular.com). In these simulations, all covalent bonds were constrained using the SHAKE algorithm [45]. The potential energy of bond-angle bending, out-of-plane bending, and dihedral angles was described with harmonic potentials or cosine series expansions [40]. The van der Waals interactions were calculated within a cut-off distance of 12.0 Å, with a smooth tapering to zero starting from 11.5 Å. The charge-charge and charge-induced dipole interactions were calculated using Ewald summation [46]. The induced dipole-induced dipole interactions were truncated at 12.0 Å. To avoid the polarization 'catastrophe', a Thole screening parameter of 0.2 was used for small separations between induced dipoles [47]. A multiple time step integration approach was used to enhance computation efficiency. A 0.5 fs time step was used for the calculation of valence interactions, including those involving bonds (SHAKE), bond angle bending, dihedral angles, and out-of-plane deformations. The short-range, non-bonded interactions (with 7.0 Å radius) were calculated every 1.5 fs, while a time step of 3.0 fs was employed for the remaining non-bonded interactions and the reciprocal part of the Ewald summation.

*System Initialization and Simulation Protocol* – Simulation cells were prepared with two different initial configurations of the molecules: (*i*) *POLAR* (*POL* – all molecules oriented along the +z direction), and (*ii*) *NONPOLAR* (*NONPOL* – equal numbers of molecules oriented along the +z and –z directions). Initially, the 384 molecules were positioned on a relatively low-density lattice with simulation cell dimensions of 150 Å in the x and y directions and 70 Å in the z direction. A 630 ps compression simulation was then conducted to achieve a mass density of ~1.0 g/cm$^3$ (comparable to typical thermotropic liquid crystal mass densities), with the z-dimension of the simulation cell fixed at 70 Å, and with a biasing potential applied to the ends of the mesogens to preserve their orientation during the initial equilibration stage. The biasing potentials were then removed and further equilibration runs 6 ns in duration and production runs in excess of 20 ns were carried out. All simulations were conducted in the NPT (isobaric, isothermal) ensemble with the z-dimension of the cell fixed and the x and y dimensions allowed to vary to maintain a constant pressure of 1 atm (NPT-XY ensemble). Each system was simulated at 110ºC, 130ºC, 150ºC, and 180ºC, temperatures spanning the $N_F$ - N phase transition, using polarizable and non-polarizable force fields. The temperature and pressure were controlled with the Nose-Hoover thermostat and barostat [48]. The same simulation protocol was used to model the cyano-terminated mesogen RM734-CN [1,2].

## SECTION S10 – ATOMISTIC SIMULATION RESULTS

*Order parameters* – An instantaneous configuration of the *POL* system at T = 130 °C is shown in *Fig. S18*, revealing that the system retains a high degree of orientational order at this temperature. To quantify nematic orientational order in this system, we measure the traceless, symmetric nematic ordering tensor $\boldsymbol{Q} = \frac{1}{N}\sum_{\alpha=1}^{N}\left(\frac{3}{2}\boldsymbol{u}_\alpha \boldsymbol{u}_\alpha - \frac{1}{2}\boldsymbol{I}\right)$, where *I* is the identity matrix, and where the sum ranges over all molecules. The scalar nematic order parameter *S* corresponds to the largest eigenvalue of the time-averaged ordering tensor $\langle Q \rangle$, and the biaxial order parameter $B$ is defined as the difference between the two smallest eigenvalues. Polar order is assessed by measuring the (vector) polar order parameter
$\mathbf{P} = \frac{1}{N}\sum_{\alpha=1}^{N} \mathbf{u}_\alpha$, from which a scalar polar order parameter $P = \langle \mathbf{P} \rangle$ can be obtained.

For the *POL* simulation of the polarizable model at 130 °C, we measure a large nematic order parameter, $S = 0.787 \pm 0.009$, and a nearly saturated polar order parameter, $\Pi = 0.924 \pm 0.003$, with negligible biaxiality ($B = 0.013 \pm 0.002$). Moreover, the polar order parameter $\mathbf{P}$ is co-linear with $\mathbf{n}$, the principal eigenvector of $\langle Q \rangle$. Given that there appear to be no long-range positional correlations (as shown below), the simulated state appears to be a uniaxial ferroelectric nematic ($N_F$) phase.

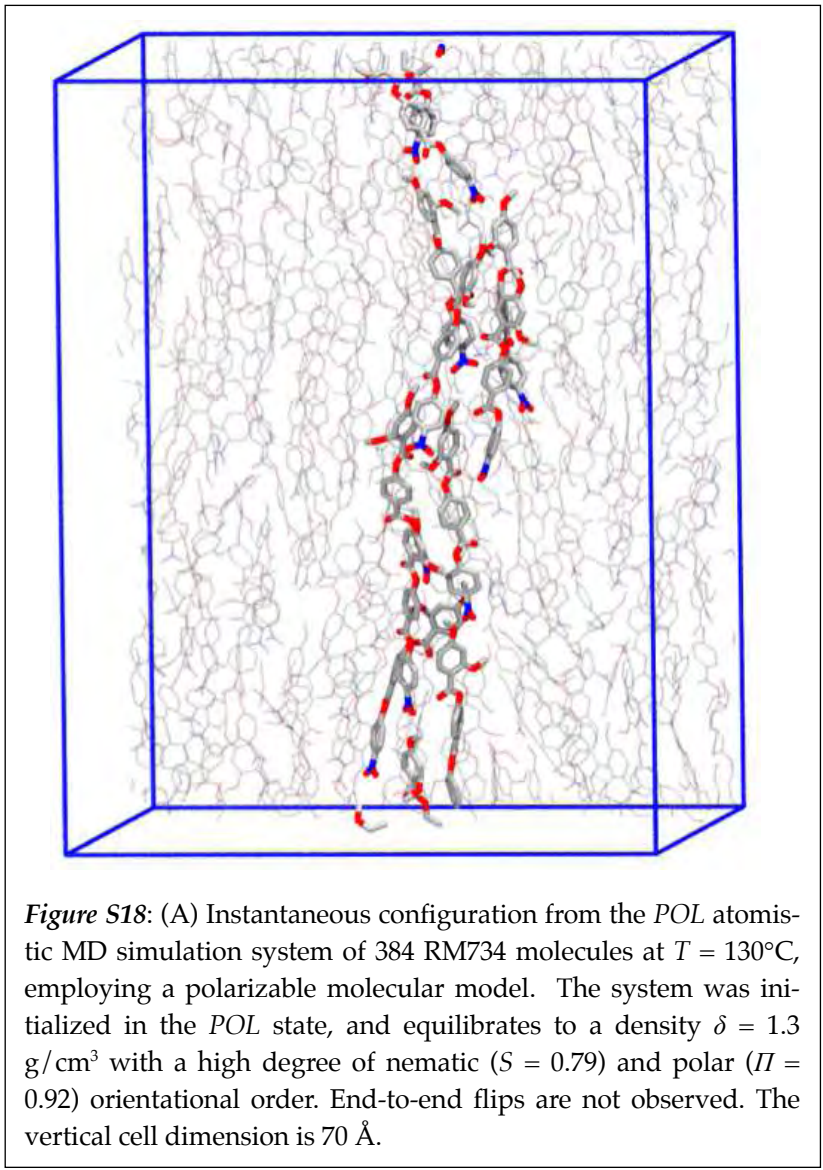

*Figure S18*: (A) Instantaneous configuration from the *POL* atomistic MD simulation system of 384 RM734 molecules at $T = 130°C$, employing a polarizable molecular model. The system was initialized in the *POL* state, and equilibrates to a density $\delta = 1.3$ g/cm³ with a high degree of nematic ($S = 0.79$) and polar ($\Pi = 0.92$) orientational order. End-to-end flips are not observed. The vertical cell dimension is 70 Å.

It is interesting to compare these results with those from the *NONPOLAR* simulation under the same conditions ($T = 130$ °C, polarizable molecular model). The nematic order parameter in this case is $S = 0.78 \pm 0.02$, quite similar to that of the polar system, while the polar and biaxial order parameters are small ($P = 0.013 \pm 0.004$, $B = 0.028 \pm 0.003$), as expected for a conventional uniaxial nematic (N) state. The fact that the magnitude of S is nearly the same in the *POL* and *NONPOL* states is generally consistent with the experimental observation that the birefringence does not change significantly through the N–$N_F$ transition. The simulated mass density at T = 130°C is $\delta = 1.3$ g/cm³.

*Ferroelectric polarization density* – The measured spontaneous ferroelectric polarization density $P$ in the $N_F$ phase of RM734 is large, increasing with decreasing temperature below the N–$N_F$ transition to a saturation value of around $P = 6$ μC/cm² (*Fig. 3*). This implies a high degree of polar order in the $N_F$ phase, that can be further quantified by comparison with the polarization $P_S$ computed by simulation.

As noted above, RM734 has a large electric dipole moment $p = 11.4$ D, as determined from quantum chemistry calculations at the B3LYP/6-31G* level of theory (*Fig. S19*). Higher-level quantum chemistry calculations were used to assign site charges to atom sites and lone pair electron sites in the molecular mechanics model used in our atomistic simulations. The resulting static site charges, shown in *Fig. S20*, are also consistent with a molecular dipole moment of around 11 D (computed using polarization $\boldsymbol{p} = \int \boldsymbol{r} \rho(\boldsymbol{r}) d\boldsymbol{r} = \sum_{i=1}^{n} \boldsymbol{r}_i q_i$, where $r_i$ and $q_i$ are the site positions and charges, and the sum ranges over all $n$ sites in the molecule). Note that the dipole moment has a weak dependence on molecular conformation, and that molecules sample an ensemble of low-energy conformations over the course of a simulation. For the *POL* simulation of the polarizable model at 130°C, we measure an average static molecular dipole moment (from static site charges) of magnitude $\langle p_{static} \rangle$ = 11.24 ± 0.01 D. For the polarizable models, there is also an induced molecular dipole moment component, which has an average magnitude of $\langle p_{induced} \rangle$ = 1.46 ± 0.02 D, but has a nearly isotropic orientational distribution, so the average magnitude of the total molecular dipole moment (the sum of static and induced contributions) is nearly equal to the static contribution, $\langle p_{total} \rangle$ = 11.20 ± 0.01 D. The fact that the induced molecular dipole has a nearly isotropic orientational distribution is a consequence of the boundary conditions, which ensure that the average electric field is zero at any point in th e system (there is no bound charge at the surface of the system, so the depolarization field vanishes), so the average magnitude of the induced dipole moment vector is close to zero, $|\langle \boldsymbol{p}_{induced} \rangle|$ = 0.053 ± 0.009 D.

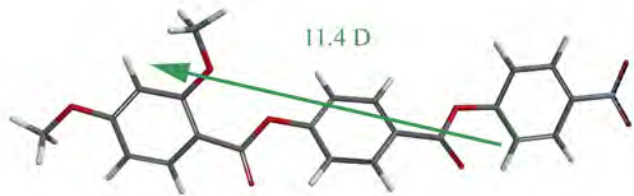

*Figure S19*: Geometry-optimized structure of RM734 computed at the B3LYP/6-31G* level of theory, showing the orientation of the 11.4 D molecular dipole moment (green arrow) for this specific molecular conformation. Other low-energy conformations have comparable dipole moments.

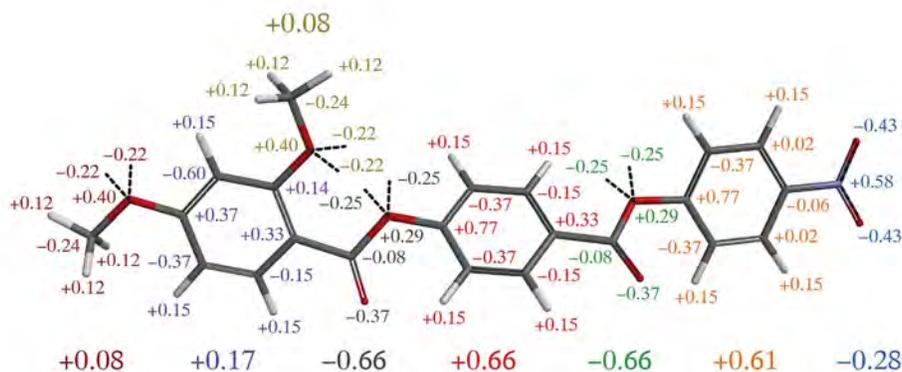

*Figure S20*: Static site charge distribution used in the atomistic simulations. The overall charges of specific functional groups, indicated in large type, show an alternation of group charges along the length of the molecule. The dashed lines correspond to lone-pair electrons.

We can gain further insight by resolving the total ferrolectric polarization density into contributions from specific dipolar groups. To accomplish this, we employ a unique decomposition of charges into elementary charge-neutral dipolar groups (bonds and rings), as shown in *Fig. S21*. Group dipoles can be further aggregated into functional groups, which are indicated by color coding in *Fig. S21*, together with the dipole moments and average contributions to ferroelectric polarization density $P_S$ of specific functional groups. The terminal nitro group and the ring to which it is attached are highly dipolar, and make a dom-inant (~ 64%) contribution to $P_S$. Four functional groups (nitro, nitro ring, central ring, and terminal methoxy) account for ~ 90% of the molecular polarization density.

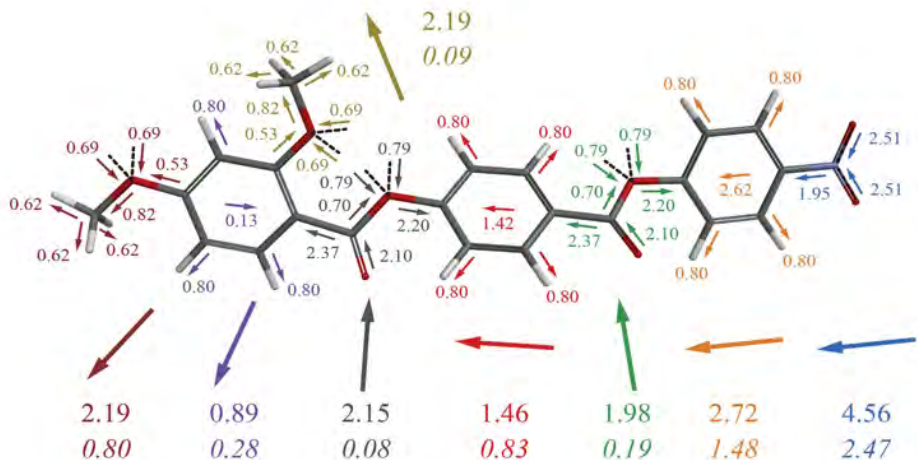

*Figure S21*: Decomposition of the static site charge distribution into group dipole contributions. Irreducible bond and ring dipole moments are shown as small arrows, where the numerical value is the dipole moment in Debye (D). The dipole moments of specific functional groups are also indicated (large arrows and large, non-italiced text). The numbers in italics are the average contributions of specific functional groups to the computed ferroelectric polarization density $P_S$, in units of $\mu C/cm^2$. The nitro group and the ring to which it is attached (the nitro ring) have the largest dipole moments, and together contribute 64% of the total polarization density. Four functional groups (nitro, nitro ring, central ring, and terminal methoxy) contribute 90% of the total polarization density. The ester groups and lateral methoxy possess substantial lateral dipole moments, which may contribute to intermolecular association.

We have calculated the average ferroelectric polarization density of the maximally polar equilibrium state of the *POL* simulation at 130°C from $\langle \mathbf{P}_S \rangle = \langle \frac{1}{V} \sum_{\alpha=1}^{N} \mathbf{p}_\alpha \rangle$, where $V$ is the system volume. For the *POLAR* simulation of the polarizable model, the polarization density magnitude $P_S = |\langle \mathbf{P}_S \rangle| = 6.17 \pm 0.01\ \mu C/cm^2$, where only $0.13 \pm 0.03\ \mu C/cm^2$ is due to the induced polarization. This calculated $P_S$ is in quantitative agreement with the saturation polarization density $P$ measured experimentally (***Fig. 3***). This value is also nearly the same as that obtained from the simple estimate given in the text assuming 100% polar order, as might be expected given the high value of the *POL* system polar order parameter given above, $\Pi = 0.924$ at $T = 110°$C. The remanant orientational disorder is due to small orientation fluctuations about $\mathbf{n}$, basically all that is allowed in the *POL* system. Consequently, $\Pi$ depends only weakly on $T$, with the MD giving $P_S = 6.170 \pm 0.008\ \mu C/cm^2$ at $T = 130°$C and $P_S = 6.368 \pm 0.002\ \mu C/cm^2$ at $T = 110°$C.

An important inference of this agreement with the $N_F$ phase experimental value is that at low $T$ RM734 essentially becomes the *POL* system, i.e. is a polar nematic with no molecular flips, and remnant polar disorder that is strictly short ranged (~few molecule) small angle orientation fluctuations about the director. At higher temperatures $P$ decreases in the $N_F$ phase because of the growth of longer length scale fluctuations and disordering modes. But these disappear upon cooling to the saturated state at low $T$, where the fluctuations in the $N_F$ become consistent with those allowed in the nanoscale volume and periodic boundary conditions of the *POL* simulation. On the other hand, the simulation volume is chosen to be large enough to observe the local molecular packing driven by specific intermolecular interactions, discussed next.

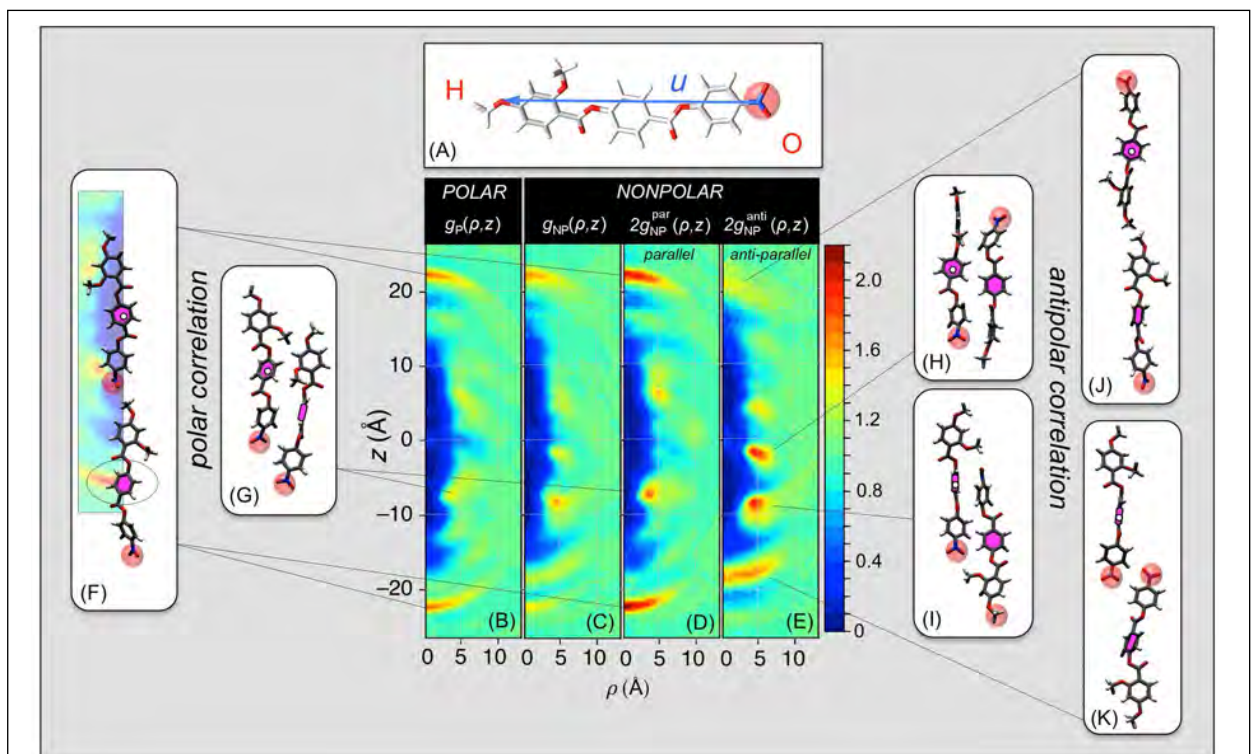

*Figure S22*: Results of atomistic molecular dynamic simulations designed specifically to explore the molecular interactions and resulting positional/orientational correlations responsible for the polar molecular ordering of RM734, shown in (**A**). A nanoscale volume containing 384 molecules is equilibrated in these simulations into two distinct LC states: a *POLAR* system with all polar molecular long axes, ***u***, along +z, and a *NONPOLAR* system with half along +z and half along -z. Equilibration of the molecular conformation and packing is readily achieved but end-to-end flips are rare so the equilibrated states remain in the limit of polar or non-polar nematic order, respectively. (**B,F,G**) The *POL* simulation shows directly the dominant pair correlations adopted by molecules that are polar ordered, in the form of conditional probability densities, $g(\rho,z)$, of molecular centers around a molecule with its center at the origin and long axis ***u*** along z. The $g(\rho,z)$ are $\varphi$-averaged to be uniaxially symmetric, reflecting the uniaxial symmetry of the N and $N_F$ phases. They exhibit a molecule-shaped, low-density region ($g(\rho,z) \sim 0$) around the origin resulting from the steric overlap exclusion of the molecules; an asymptotic constant value at large $\rho$ giving the normalized average density ($g(\rho,z) = 1$); and distinct peaks indicating preferred modes of molecular packing. This analysis reveals two principal preferred packing modes in the *POL* system: (**B,F**) polar head-to-tail association stabilized by the attraction of the terminal nitro and methoxy groups, and (**B,G**) polar side-by-side association governed by group charges along the molecule, nitro-lateral methoxy attraction, and steric interactions of the lateral methoxys. (**D,E**) The *NONPOL* system exhibits distinct correlation functions for antiparallel and parallel molecular pairs, $g_{NP}^{anti}(\rho,z)$ and $g_{NP}^{par}(\rho,z)$. (**D,H,I**) The preferred antiparallel packing gives strong side-by-side correlations, governed by group charges along the molecule; and (**D,J,K**) weaker antipolar nitro-nitro end-to-end association. (**D,F,G**) The parallel correlations in the *NONPOL* system are the most relevant to the stability of polar order in the $N_F$ phase, as they are determined by the inherent tendency of the molecular interactions for polar ordering in the presence of enforced polar disorder. Comparison of (**B**) and (**D**) shows identical preferred modes of parallel association in the two systems, with the *POL* system correlations being even stronger in the *NONPOL* system. This is clear evidence that the polar packing motifs giving the correlation functions (**B**) and (**D**), exemplified by the sample *POL* configurations (**F**) and (**G**), stabilize the polar order of the $N_F$.

*Intermolecular correlations* – In order to make headway in understanding the roles of molecular structure and interaction in RM734 and its relation to polar ordering, we applied the atomistic simulations to probe molecular association and packing in the *POL* and *NONPOL* systems. We characterized molecular pair positional and orientational correlations by measuring several $g(\rho,z)$, the conditional densities of molecular centers about a molecule fixed with its center at the origin, with its long axis $u$ in **Fig. S22A** parallel to $z$ and pointing in the +z direction [nitro group near $(\rho,z) = (0,-10$ Å) and the methoxy group near $(\rho,z) = (0,10$ Å)]. Here the center of the molecule is defined as the midpoint of $u$ in **Fig. S22A**, and the $g(\rho,z)$ are angular averages about $z$ of density over the azimuthal orientations of the molecule at the origin. The $g(\rho,z)$ are therefore independent of azimuthal angle $\varphi$, reflecting uniaxial nematic symmetry. They all exhibit a correlation hole ($g(\rho,z) \sim 0$) around the origin and extended along $z$ for ~ 1.5 molecular lengths where other molecular centers are excluded because of steric repulsion and the strong nematic orientational ordering. The pair distributions $g_P(\rho,z)$ and $g_{NP}(\rho,z)$ for the *POL* and *NONPOL* systems, shown in **Fig. S22B**, and **S22C-E**, respectively, display a number of striking features indicating specific molecular association motifs. In the *POL* system, pronounced arc-shaped peaks observed near ($\rho = 0$ Å, $z = \pm 22$ Å) indicate a strong tendency for head-to-tail association of parallel pairs of molecules, as illustrated by the representative pair configuration shown in **Fig. S22C**. Such head-to-tail configurations are characterized by close association of positively charged H atoms in the terminal methoxy group of the molecule at the origin with negatively charged O atoms in the nitro group of its neighbor, suggesting that head-to-tail association is in large part the result of specific electrostatic interactions. Prominent *off-axis* peaks near $\rho = 5$ Å, $z = \pm 6$ Å are also observed in $g_P(\rho,z)$, and analysis of pair configurations associated with these peaks (e.g., **Fig. S22E**) reveals close association of oppositely charged atoms, including close contacts between positively charged H atoms in the lateral methoxy group and the negatively charged O atom in the terminal methoxy group, and between negatively charged O atoms in the nitro and carbonyl groups and positively charged H atoms in the phenyl rings. These observations also point to specific electrostatic interactions stabilizing polar pair configurations. In the *POLAR* simulation the lateral methoxys appear to be key to establishing the relative positioning of the side-by-side molecular associates that prefer polar ordering. The presence of a region of reduced probability ('correlation hole') in $g_P(\rho,z)$ near $\rho = 5$ Å, $z = 0$ Å shows that side-by-side configurations of parallel pairs of molecules are relatively unfavorable, presumably due to the excluded volume of their lateral methoxy groups.

The *NONPOL* system forces both antiparallel and parallel molecular pairs, which give correlation functions, $g_{NP}^{anti}(\rho,z)$ and $g_{NP}^{par}(\rho,z)$, that exhibit very strongly expressed, polarity-dependent molecular recognition. In the on-axis or nearly on-axis peaks of $g_{NP}^{anti}(\rho,z)$ in **Fig. S22G** we observe that the $z \rightarrow -z$ symmetry of $g_{NP}^{anti}(\rho,z)$ is the most strongly broken, as expected since HO-OH association will be different from OH-HO association. Thus, there is a prominent arc-shaped peak near $\rho = 4$ Å, $z = -18$ Å, arising from HO-OH antiparallel lateral association of terminal nitro groups, illustrated by the representative pair configuration in **Figs. S22G,K**. Pair configurations associated with this peak are characterized by close contacts between negatively charged nitro O atoms and positively charged phenyl H atoms adjacent to the nitro group, suggesting that these configurations are electrostatically stabilized. In contrast, the OH-HO associations between terminal methoxy groups (associated with the peak near $\rho = 0$ Å, $z = +21$ Å) are very weak (**Figs. S22G,J**).

The *NONPOL* system contribution to $g_{NP}^{anti}(\rho,z)$ from side-by-side antiparallel pairs (**Figs. S22G-I**), shows a prominent peak near $\rho = 5$ Å, $z = -8$ Å, associated with pair configurations characterized by close contacts between positively charged H atoms in the lateral methoxy group and negatively charged nitro O atoms (an example of which is shown in **Fig. S2I**). The side-by-side pair configurations associated with the peak near $\rho = 5$ Å, $z = -2$ Å similarly involve close contacts between oppositely charged O and H at-oms in the ester groups and phenyl rings. These findings provide further support for the hypothesis that electrostatic interactions between specific oppositely charged atoms play a dominant role in stabilizing the characteristic pair configurations observed in our simulations.

The *NONPOL* system parallel pair correlation function $g_{NP}^{par}(\rho,z)$ in **Fig. S22F** is very similar to the *POL* system $g_P(\rho,z)$ in **Fig. S22C**, indicating a nanosegregation of the *par* and *anti* components, a mixture of OH-OH-OH and HO-HO-HO chains with the OH-HO at their interfaces. Remarkably, the polar features of $g_P(\rho,z)$ are not only dominant in $g_{NP}^{par}(\rho,z)$ but even more pronounced than in $g_P(\rho,z)$ itself. This suggests that there are certain polar associations in the *POL* system that can reduce the overall polar order, but that can be replaced by antipolar associations in the *NONPOL* system that are more favorable for nearby polar order. In any case, the simulations show that the polar correlations in **Figs. S22B,C,F,E**, emergent in both *POL* and *NONPOL* systems, are the only packing motifs stabilizing polar order and therefore must be the principal drivers stabilizing the $N_F$ phase.

*The Plupolar Nematic* – The *POL* simulation equilibrates a state in which end-to-end flipping is kinetically arrested and the periodic boundary conditions constrain the allowed wavelengths of orientation fluctuations to $\lambda_x < 55$ Å and $\lambda_z < 70$ Å. The remnant short ranged fluctuations create the pair correlations exhibited in **Fig. S22**, which are confined to the volume $\rho < 10$ Å and $z < 30$ Å about the origin, molecular neighbor separation scales, well within the dimensions of the simulation box. These conditions create a *"plupolar" (plus quam polar* [49]) equilibrium state of constrained polar ordering yielding the simulated $P_S$ values in **Fig. 3** (open circles). Comparing these values with the RM734 data shows that, on the one hand, in the *plupolar* state the fluctuations that lead to the phase transition are clearly suppressed, while the remnant short range fluctuations give a $P_S$ value exhibiting only a weak dependence on temperature. On the other hand, this $P_S$ gives a good account of the polarization density of the $N_F$ at low temperature, evidence that at low $T$ the $N_F$ phase approaches some comparable *plupolar*-like condition of having only short-range fluctuations, and that the simulated $g(\rho,z)$ are characterizing their remnant correlations. This state may be glassy, if the strong $T$-dependence of the viscosity is any indicator.